\documentclass{article}
\usepackage[nonatbib, preprint]{neurips_2020}
\usepackage{url}
\usepackage{latexsym,amssymb}
\usepackage[cmex10]{amsmath}

\usepackage{tikz}

\usepackage{caption}
\usepackage{subcaption}
\usepackage{multirow}
\usepackage{booktabs}
\usepackage{bm}
\usepackage{array}
\newcolumntype{C}[1]{>{\centering\let\newline\\\arraybackslash\hspace{0pt}}m{#1}}

\usepackage{pifont}

\newcommand\T{\rule{0pt}{2.9ex}}       
\newcommand\B{\rule[-1.2ex]{0pt}{0pt}} 

\title{Understanding the Performance of Knowledge Graph Embeddings in Drug Discovery}

\author{
    Stephen Bonner\textsuperscript{1},\enskip Ian P Barrett\textsuperscript{1},\enskip Cheng Ye\textsuperscript{1},\enskip Rowan Swiers\textsuperscript{1} \\
\rule[10pt]{0pt}{0pt}
\textbf{Ola Engkvist\textsuperscript{2}, Charles Tapley Hoyt\textsuperscript{3}, William L Hamilton\textsuperscript{4,5}} \\
\rule[15pt]{0pt}{0pt}
{\normalsize \textsuperscript{1}Data Sciences and Quantitative Biology, Discovery Sciences, R\&D, AstraZeneca, Cambridge, UK}\\
{\normalsize \textsuperscript{2}Molecular AI, Discovery Sciences, R\&D, AstraZeneca, Gothenburg, Sweden}\\
{\normalsize \textsuperscript{3}Laboratory of Systems Pharmacology, Harvard Medical School, Boston, USA}\\
{\normalsize \textsuperscript{4}School of Computer Science, McGill University, Montreal, Canada} \\
{\, \normalsize \textsuperscript{5}Mila - Quebec AI Institute, Montreal, Canada}\\
\rule[15pt]{0pt}{0pt}
}

\begin{document}
\maketitle

\begin{abstract}

    Knowledge Graphs (KG) and associated Knowledge Graph Embedding (KGE) models have recently begun to be explored in the context of drug discovery and have the potential to assist in key challenges such as target identification. In the drug discovery domain, KGs can be employed as part of a process which can result in lab-based experiments being performed, or impact on other decisions, incurring significant time and financial costs and most importantly, ultimately influencing patient healthcare. For KGE models to have impact in this domain, a better understanding of not only of performance, but also the various factors which determine it, is required.

    In this study we investigate, over the course of many thousands of experiments, the predictive performance of five KGE models on two public drug discovery-oriented KGs. Our goal is not to focus on the best overall model or configuration, instead we take a deeper look at how performance can be affected by changes in the training setup, choice of hyperparameters, model parameter initialisation seed and different splits of the datasets. Our results highlight that these factors have significant impact on performance and can even affect the ranking of models. Indeed these factors should be reported along with model architectures to ensure complete reproducibility and fair comparisons of future work, and we argue this is critical for the acceptance of use, and impact of KGEs in a biomedical setting.

\end{abstract}

\section{Introduction}\label{sec:intro}

The task of discovering effective and safe drugs is a complex and interdisciplinary one, with many drugs failing in clinical trials before being able to help patients~\cite{morgan2018impact}. Thus, the field is looking to leverage the large quantities of available data and information, much of which is inherently interconnected, to help improve the chances of a successful drug making it to market. Consequently, over recent years, an increasing number of Knowledge Graphs (KG) suitable for use within the drug discovery domain have been created~\cite{himmelstein2017systematic, walsh2020biokg}, where drugs, genes and diseases are used as entities, with the interactions between them captured as relations. Several fundamental tasks within drug discovery can then be thought of as predicting the missing links between these entities -- for example, drug repurposing can be considered as predicting missing links between drug and disease entities~\cite{malas2019drug} or target discovery as identifying missing links between genes and diseases~\cite{paliwal2020preclinical}.

Naturally the family of Knowledge Graph Embedding (KGE) models (approaches which learn low dimensional representation of entities and relations which are trained to predict the plausibility of a given triple) have begun to be employed for these tasks. Perhaps unlike other domains, the predictions from such models are part of processes which can result in physical real-world experimentation being performed, and ultimately even clinical trials being undertaken -- both with significant financial, regulatory and time costs associated, and more importantly impacting on the efforts to improve patient health. Therefore, there is an obvious need to ensure not only accurate predictions are being made but also that there is clear understanding of the various factors that can affect model predictive performance so that such predictions can be used effectively to derive impact and value. One other interesting aspect to consider is that these KGE models are typically not designed or tested against drug discovery datasets, so there is limited understanding about how they should be expected to perform in such cases. Indeed, recent work~\cite{liu2021neural} has shown that biomedical knowledge graphs have markedly different topological structure to typical KG benchmark datasets such as FB15K-237~\cite{toutanova2015observed} or WN18RR~\cite{dettmers2018convolutional}, displaying much greater average connectivity. Thus adding motivation for understanding how KGE models perform on such datasets.

In this paper, we perform a detailed experimental study of various factors that can affect KGE model performance, using five model architectures (ComplEx~\cite{trouillon2016complex}, DistMult~\cite{yang2015embedding}, RotatE~\cite{sun2019rotate}, TransE~\cite{bordes2013translating} and TransH~\cite{wang2014knowledge}) and two real-world drug discovery oriented KGs (Hetionet~\cite{himmelstein2017systematic} and BioKG~\cite{walsh2020biokg}) with the goal of aiding better understanding, evaluation practices and reproducibility in the domain. The factors we investigate are the training setup of the model, the impact of changes in model hyperparameters, how model performance can be affected by both different random initialisations and changes in the train/test dataset splits and assessing performance on a domain specific task. All experiments are performed under a unified and consistent evaluation framework, on public data sources, using known best practices to ensure fair and reproducible comparisons. Additionally we release code to replicate our results\footnote{\url{https://github.com/AstraZeneca/kgem-in-drug-discovery}}.

Lastly, we note that the contribution of this work is not to achieve state-of-the-art results on a given dataset or even to decide upon the definitive choices for the various factors we are investigating, rather it is to highlight how these can affect overall predictive performance and to encourage further research and attention on such foundational topics. Indeed, to the best of our knowledge, this is the first work to specifically focus on KGE model performance factors in the drug discovery domain.
\section{Related Work}\label{sec:litreview}

\subsection{Knowledge Graph Embeddings}\label{ssec:kgs}

\textbf{Knowledge Graphs.} A Knowledge Graph is a heterogeneous, multi-relation and directed graph, containing information about a set of entities \(\mathcal{E}\) and a set of relationships between them \(\mathcal{R}\), defined as \(\mathcal{K} \subseteq \mathcal{E} \times \mathcal{R} \times \mathcal{E}\)~\cite{zhang2019heterogeneous}. A Knowledge Graph is often considered as a series of triples \((h,r,t) \in \mathcal{K}\), where \( h,t \in \mathcal{E}\) are the head and tail entities connected via the relationship \( r \in \mathcal{R}\). A hypothetical triple from a drug discovery knowledge graph could be \(( Drug_1, treats, Disease_2) \), where the entities \(Drug_1\) and \(Disease_2\) are connected via the relationship \(treats\). In many real-world knowledge graphs, the set of triples is known to be noisy and incomplete~\cite{ali2020bringing}. Thus numerous techniques have emerged which attempt to complete the missing knowledge based on the existing data in \(\mathcal{K}\) through multi-relation link prediction~\cite{ruffinelli2019you}. Such techniques consider the partial triple \(( Drug_1, treats, ?) \) and attempt to predict the correct tail entity, or be given \(( ?, treats, Disease_2) \) and attempt to predict the correct head entity.

\textbf{Knowledge Graph Embeddings.} A growing number of approaches have been proposed in the literature which attempt to perform this knowledge graph completion task. In this work we focus upon the family of Knowledge Graph Embedding (KGE) techniques~\cite{ji2021survey, wang2017knowledge}. Typically, a KGE model learns a low-dimensional representation of each entity and relation in the graph. These embeddings are combined in various ways to produce a scalar value representing a measure of how likely that triple is to be true, with a larger score typically implying a more plausible triple~\cite{ali2020bringing}. More concretely, a model \(f : \mathcal{E} \times \mathcal{R} \times \mathcal{E} \rightarrow \mathbb{R} \), calculates a scalar value representing the plausibility for each potential triple \((h,r,t) \in \mathcal{K}\). For KGE approaches, \(f\) is typically a learned model which operates only with the embeddings of the elements in the triples, \(f( \mathbf{h}, \mathbf{r} , \mathbf{t} )\), where \(\mathbf{h}, \mathbf{t} \in \mathbb{R}^k \) and \(\mathbf{r} \in \mathbb{R}^j \). The values of \(k\) and \(j\) represent the dimension of the entity and relation embedding respectively\footnote{In practice, these are often set as the same value.}.

\subsection{Understanding Knowledge Graph Embedding Model Performance}

Over recent years, there has been increasing interest in machine learning with graph structured data, with approaches created for homogeneous graph embeddings~\cite{zhang2018network}, graph-specific neural models~\cite{hamilton2017representation} and knowledge graph embedding~\cite{ji2021survey}. Whilst there have been numerous new model architectures proposed in the literature, there has been less work performed on understanding how these models are affected by the rest of the choices made in the machine learning pipeline, for example, how robust they are across hyperparameter values and model initialisations, or how performance changes across dataset splits. However, the work that does exist demonstrates some interesting observations.

For example, several graph neural network approaches were compared under a fair evaluation procedure~\cite{shchur2018pitfalls}, which showed that a change in train/test split would drastically alter the ranking of the models and that simpler baseline approaches, with correctly tuned hyperparameters, could outperform more complex models. The performance of different graph neural networks for graph-level classification has also been compared~\cite{errica2020fair}, with results showing that baselines approaches not using the graph structure can outperform those that do. The need for consistent, rigorous and reproducible benchmarks for graph machine learning is also an area of increasing research interest~\cite{dwivedi2020benchmarking, hu2020open}.

\textbf{Evaluation of Knowledge Graph Embeddings}
A study comparing seven different knowledge graph embedding techniques under a consistent evaluation framework on the non-biomedical benchmark datasets FB15K-237~\cite{toutanova2015observed} and WNRR~\cite{dettmers2018convolutional} has been performed~\cite{ruffinelli2019you}. The authors observe that as new models are introduced, they are often accompanied with new training regimes or objective functions, making assessing the value of the new model architecture alone challenging. They undertake a detailed comparison across combinations of models, training paradigms and hyperparameters, using a Bayesian search approach. They find that earlier and comparatively simpler models, are very competitive when trained using modern techniques~\cite{ruffinelli2019you}. However, the study did not consider how model initialisation or dataset splits can affect performance.

In a similar study, 19 knowledge graph embedding approaches, implemented in the PyKEEN framework~\cite{ali2020pykeen}, are compared across eight different benchmark datasets~\cite{ali2020bringing}. One of the aims of the study was to investigate whether original published results could be reproduced, a task they found challenging. Additionally they perform detailed experiments over models and training paradigm combinations, searching over the hyperparameter space for a maximum of 24 hours or 100 training repeats. Again they find that suitably tuned simple models can out-perform complex ones. The study does not consider drug discovery datasets specifically and does not assess how models perform across model seeds or dataset splits.

\textbf{Biomedical Domain Specific Evaluations}
The use of various homogeneous graph embedding techniques has been assessed across a range of biomedical tasks such as drug-drug and protein-protein interactions~\cite{yue2020graph}.
Whilst not exploring knowledge graph embedding techniques, the work explores how various hyperparameters affect predictive performance.
They explore random walk and neural network based techniques including DeepWalk~\cite{perozzi2014deepwalk} and Graph Convolution based auto-encoders~\cite{kipf2016variational}, using various task specific homogeneous graphs.
An additional review compares both graph and knowledge graph specific approaches and their use in the biomedical world, however no experimental comparisons are made between the different approaches~\cite{su2020network}.

The performance of five knowledge graph embedding approaches (TransE, ComplEx, DistMult, SimplE and RotatE) have been compared on a knowledge graph constructed from the SNOMED resource~\cite{chang2020benchmark}.
The models are assessed on the tasks of link prediction, visualisation and entity classification, with a limited grid-search being performed to choose the hyperparameters.

Work has assessed the performance of knowledge graph embedding approaches for tasks within the drug discovery domain such as predicting drug-target interactions~\cite{mohamed2020biological}.
A different, often task-specific, graph is used for each of these experimental setups with much of the data being taken from the BioSNAP repository~\cite{biosnapnets}.
However often these graphs are less complex than resources like Hetionet, with a typically limited number of entity and relationship types being present.
Results use k-fold cross validation to assess performance variability over dataset splits, with a grid-search over a range of hyperparameters also being performed.

The effect of different data splitting strategies for predicting drug-drug interaction using graph-based methods (including TransE and TransD) has been investigated~\cite{celebi2019evaluation}.
The work argues that realistic data splits should be used in order to avoid over-optimistic results, with several domain specific and time-based splits being assessed.
Additionally the work claims that the tuning of various hyperparameters had little impact on the overall model performance.

\subsection{Knowledge Graphs in Drug Discovery}

Recently, approaches exploiting knowledge graphs are being leveraged within the drug discovery domain to solve key tasks~\cite{bonner2021review, gaudelet2021utilizing}. In a drug discovery knowledge graph, entities often represent key elements such as genes, disease or drugs, whilst the relations between them capture interactions. Many important tasks in drug discovery can then be considered as predicting missing links between these entities. For example, performing drug target identification, the process of finding genes involved in the mechanism of a given disease, has been addressed as link prediction between gene and disease entities using the ComplEx model on a drug discovery graph~\cite{paliwal2020preclinical}.

There are increasing numbers of public knowledge graphs suitable for use in drug discovery~\cite{bonner2021review}. One of the first such graphs was Hetionet~\cite{himmelstein2017systematic}, originally created for drug purposing through the use of knowledge graph-based approaches. Since its introduction, other datasets have been released including the Drug Repurposing Knowledge Graph (DRKG)~\cite{drkg2020}, OpenBioLink~\cite{breit2020openbiolink} and BioKG~\cite{walsh2020biokg}.
\section{Experimental Setup}

In this section, we give an overview of the models, datasets and evaluation protocol used for our experimentation.

\subsection{Models}

As detailed in Section~\ref{ssec:kgs}, many knowledge graph specific embedding models have been introduced, with the primary differentiator between them being how they score the plausibility of a given triple. Here we briefly detail the models utilised in this study, but interested readers are referred to larger reviews for context and comparisons with other approaches~\cite{ji2021survey, rossi2021knowledge, wang2017knowledge}. The models we have selected are popular approaches from the literature, cover a range of different methodologies and have begun to be explored in the context of drug discovery~\cite{paliwal2020preclinical, zheng2020pharmkg}.

\textbf{DistMult.} A simplification of the earlier RESCAL model, DistMult uses a vector for each relation type (represented as a diagonal square matrix to greatly reduce the number of required parameters) but is limited to learning only symmetric relations~\cite{yang2015embedding}. The score function used by DistMult to evaluate each triple is thus as follows:
\[ f(h,r,t) =  \mathbf{h}^\top \mathrm{diag}(\mathbf{r}) \mathbf{t} \]

\textbf{ComplEx.} To help overcome the ability to only learn symmetric relations, an extension of DistMult called ComplEx has been introduced~\cite{trouillon2016complex}. The entity and relation embeddings for the ComplEx model are not real valued, unlike many other approaches, instead they are complex valued such that \( \mathbf{h}, \mathbf{r}, \mathbf{t} \in \mathbb{C}^k \). The score function for ComplEx then becomes:
\[ f(h,r,t) =  Re(\mathbf{h} \odot  \mathbf{r} \odot  \mathbf{t}), \]
where \(\odot\) is the Hadamard product and \(Re()\) is real value only from the complex number.

\textbf{TransE.} One of the first models to use translational distance to learn embeddings such that relations are used to translate in latent space is TransE~\cite{bordes2013translating}. In TransE, the relation embedding is added to the head entity such that the result lies close to the tail embedding, where the score function can be thought of as:
\[ f(h,r,t) =  -|| \mathbf{h} + \mathbf{r} - \mathbf{t}||_F, \]
where \(F\) is typically either the l1 or l2 norm. One well known limitation of TransE is that is cannot correctly account for one-to-many, many-to-one or many-to-many relations being present in a knowledge graph~\cite{rossi2021knowledge}.

\textbf{TransH.} To help address the issues of TransE, another translational distance based model entitled TransH has been introduced~\cite{wang2014knowledge}. TransH allows for entity embeddings to be given a different context depending upon the relation used in certain triple. This is achieved by modelling each relation \(r\)  as a hyperplane, with the head and tail entity embeddings first being transformed using a normal vector of the hyperplane \(\mathbf{w}_r \in \mathbb{R}^k\) as follows:
\[ \mathbf{h}_r =  \mathbf{h} - \mathbf{w}_r^\top \mathbf{h} \mathbf{w}_r ,\]
\[ \mathbf{t}_r =  \mathbf{t} - \mathbf{w}_r^\top \mathbf{t} \mathbf{w}_r .\]

The score function is similar to the one used with TransE:
\[ f(h,r,t) =  -|| \mathbf{h}_r + \mathbf{d}_r - \mathbf{t}_r||_2^2, \]
where \(\mathbf{d}_r\) is a vector that lies in relations \(r\) hyperplane.

\textbf{RotatE.} Combining ideas from different existing models, RotatE uses complex valued embeddings for entities and relations and works such that relations rotate head to tail entities~\cite{sun2019rotate}. Given complex embeddings for head, relation and tail \( \mathbf{h}, \mathbf{r}, \mathbf{t} \in \mathbb{C}^k \) (RotatE limits the complex elements of the relation embeddings to have a modulus of one: \(|\mathbf{r}|=1\)) the score function then becomes:
\[ f(h,r,t) =  -|| \mathbf{h} \odot \mathbf{r} - \mathbf{t}||, \]
with \(\odot\) again being the Hadamard product. This allows RotatE to deal with symmetry/antisymmetry, inversion, and composition relation types~\cite{sun2019rotate}.

\subsection{Datasets}

Throughout this work we employ two publicly available knowledge graphs suitable for use within the drug discovery domain - Hetionet~\cite{himmelstein2017systematic} and BioKG~\cite{walsh2020biokg}, which are detailed in Table~\ref{tab:datasets}.

\begin{table}[ht!]
    \centering
    \small
    \begin{tabular}{p{0.2\textwidth}  C{0.1\textwidth}  C{0.1\textwidth} C{0.1\textwidth} C{0.1\textwidth}}
        \toprule
        \textbf{KG Dataset}                            & \(|\mathcal{E}|\) & \(|\mathcal{K}|\) & \(\mathcal{E}\) \textbf{Types} & \(\mathcal{R}\) \textbf{Types} \\
        \midrule \midrule
        Hetionet v1.0~\cite{himmelstein2017systematic} & 47K               & 2.2M              & 11                             & 24                             \\
        BioKG v1.0~\cite{walsh2020biokg}               & 105K              & 2M                & 10                             & 17                             \\
        \bottomrule
    \end{tabular}
    \vspace{5pt}
    \caption{Biomedical Knowledge Graphs used in this study.}\label{tab:datasets}
\end{table}

Both these datasets contain information about the key elements of drug discovery: genes, diseases and drugs, whilst capturing the interactions between them. Both are constructed from various high-quality public sources of biological and chemical information~\cite{bonner2021review}, with the major differences being in the complexity of the relationships captured (BioKG only uses a single relationship type between two entities, whereas Hetionet has up to three).

It has been observed that biomedical knowledge graphs can exhibit a different topological structure than the benchmark datasets against which the models are typically tested~\cite{liu2021neural}. For example, Hetionet has a higher average degree than datasets like FB15K-237~\cite{toutanova2015observed} and WN18RR~\cite{dettmers2018convolutional}. However it remains unknown how this impacts KGE model performance and we leave a detailed comparison between knowledge graphs from the biomedical and other domains for further work.

\subsection{Evaluation Protocol}

The evaluation of KGE models is typically performed by measuring how likely the model ranks a holdout set of triples from the original graph. However there are a series of choices one can make in this evaluation process which can drastically alter the results, thus ultimately making direct comparisons between published results challenging~\cite{ali2020bringing}. Hence the evaluation protocol is one of the crucial aspects for reproducibility, as the choices made can have a large impact on comparative performance. Here we describe our own evaluation procedure in full which closely follows the one established in~\cite{ali2020bringing}.

Given a set of test triples \(\mathcal{K}_{test} \subseteq \mathcal{K}\), we perform \emph{both} left and right side evaluation where the head and tail entities are removed in turn and the model is evaluated based on how well these missing entities can be predicted given the partial triple. To do this, for each triple in the test set \((h,r,t) \in \mathcal{K}_{test}\), two corrupted sets are constructed: the set where the head entity has been corrupted with every possible entity \( \mathcal{H}^\prime = \{(h^\prime,r,t) \mid h^\prime \in \mathcal{E} \} \) and likewise for the tail entity \( \mathcal{T}^\prime = \{(h,r,t^\prime) \mid t^\prime \in \mathcal{E} \} \). The goal then is to have the original true triple given a higher score by the model than these corrupted triples. One decision that needs to be taken is if any true triples, those that are already part of \(\mathcal{K}\), in the corrupted sets are removed before scoring. Following prior work~\cite{bordes2013translating, ali2020bringing}, we use the filtered evaluation setting where we remove any corrupted triple which is already in the graph, as their presence can skew the results.

It is possible that two or more triples in the test set are given the same score by the model when all are being ranked and how this situation is handled can also affect the results. One can assume the extremes, where the true triple is assumed to be at the start or end of the ranked list. For this work we present the mean of the rank using these two assumptions.

\textbf{Metrics.} We employ commonly used knowledge graph performance metrics including Mean Reciprocal Rank (MRR) and Hits@k (see~\cite{ali2020bringing} for definitions). Additionally we use the recently introduced Adjusted Mean Rank (AMR)~\cite{berrendorf2020on} owing to its ability to allow comparison between graphs of different sizes. However we would like to highlight that using metrics alone, especially for use cases like drug discovery where model predictions will often result in real-world lab-based experiments being performed, perhaps should not be the sole way in which models are judged.

\subsection{Implementation Details}

All work has been performed using the PyKEEN framework~\cite{ali2020pykeen}, a python library for knowledge graph embeddings built on top of PyTorch~\cite{paszke2019pytorch}. Additionally we use the Optuna library to perform the hyperparameter optimisation~\cite{akiba2019optuna}. All experiments were performed on machines with Intel(R) Xeon(R) Gold 5218 CPUs and NVIDIA(R) V100 32GB GPUs. Additionally, we kept the software environment consistent throughout all experimentation using python 3.8, CUDA 10.1, PyTorch 1.7, Optuna 2.3 and PyKEEN 1.0.6.
\section{Results}\label{sec:results}

We now present the results of our experimental evaluation. All the results presented are taken from a random 10\% holdout set of test triples, unseen during the training process. Unless otherwise stated, these test triples comprise all entity and relation types. Throughout we will make use of the default set of training setup choices and hyperparameters detailed in Table~\ref{tab:default_parms}.

\begin{table}[h!]
    \small
    \centering

    \begin{tabular}{l c c c c c }
        \toprule
        \textbf{Parameter} & \multicolumn{5}{c}{\textbf{Value By Approach}} \T\B                                          \\
        \midrule \midrule
                           & ComplEx                                              & DistMult & RotatE & TransE & TransH\B \\
        \cline{2-6}

        Embedding Dim      & 50                                                   & 50       & 200    & 50     & 50\T     \\
        Num Epochs         & \multicolumn{5}{c}{500}                                                                      \\
        Learning Rate      & \multicolumn{5}{c}{0.02}                                                                     \\
        Num Negatives      & \multicolumn{5}{c}{1}                                                                        \\
        \midrule
        Optimiser          & \multicolumn{5}{c}{Adagrad}                                                                  \\
        Inverse Relations  & \multicolumn{5}{c}{False}                                                                    \\
        Loss Function      & \multicolumn{5}{c}{Margin Ranking Loss (Margin 1.0)}                                         \\
        \bottomrule
    \end{tabular}
    \vspace{5pt}
    \caption{Default model training setups and hyperparameters.}\label{tab:default_parms}
\end{table}

\subsection{Training Setup Study}

We begin by assessing the impact of various categorial choices about the training setup of the models, evaluating several common options. Specifically, we vary the optimiser (from a choice of Stochastic Gradient Descent (SGD), Adam and AdaGrad), training objective function (from a choice of Binary Cross Entropy (BCEL), Softplus (SPL), Margin Ranking (MRL) and the self adversarial loss (called NSSA henceforth) from~\cite{sun2019rotate}) and with/without adding inverse relationships into the graph (the process of adding a copy of each triple during training with an inverse relation~\cite{kazemi2018simple}). Throughout all of these runs, the model initialisation seed, dataset split and hyperparameters were kept constant and are detailed in Table~\ref{tab:default_parms}.

\begin{figure}[!ht]
    \centering
    \begin{subfigure}[b]{0.48\textwidth}
        \centering
        \includegraphics[width=0.96\textwidth]{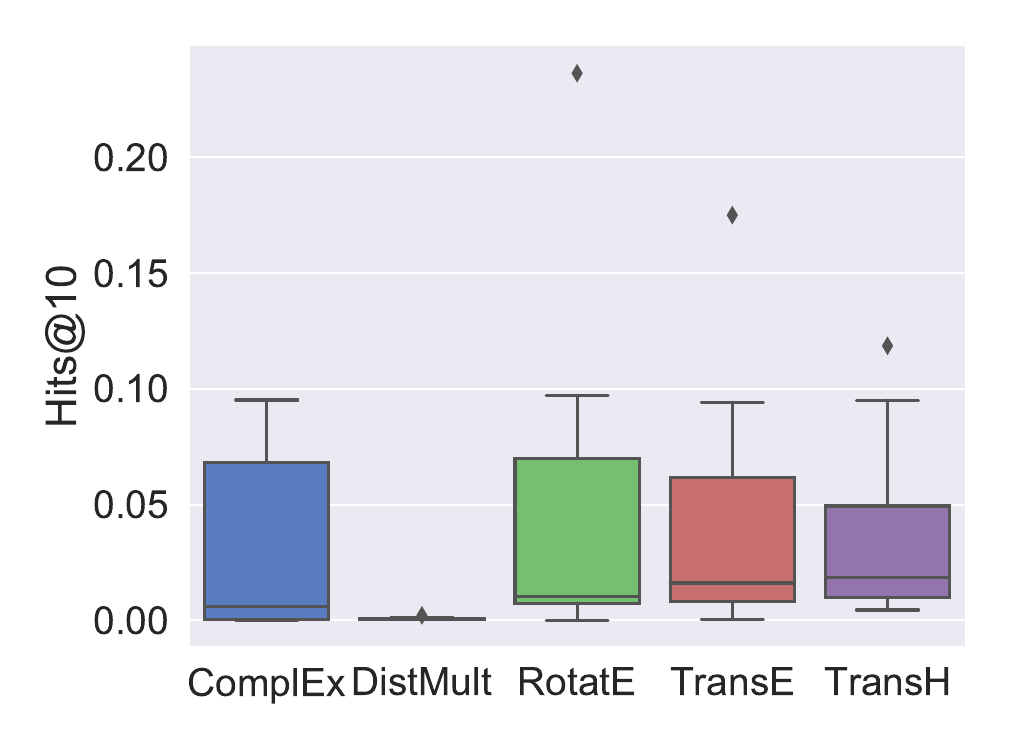}
        \caption{Hetionet}\label{fig:ab:hetnet}
    \end{subfigure}
    \begin{subfigure}[b]{0.48\textwidth}
        \centering
        \includegraphics[width=0.99\textwidth]{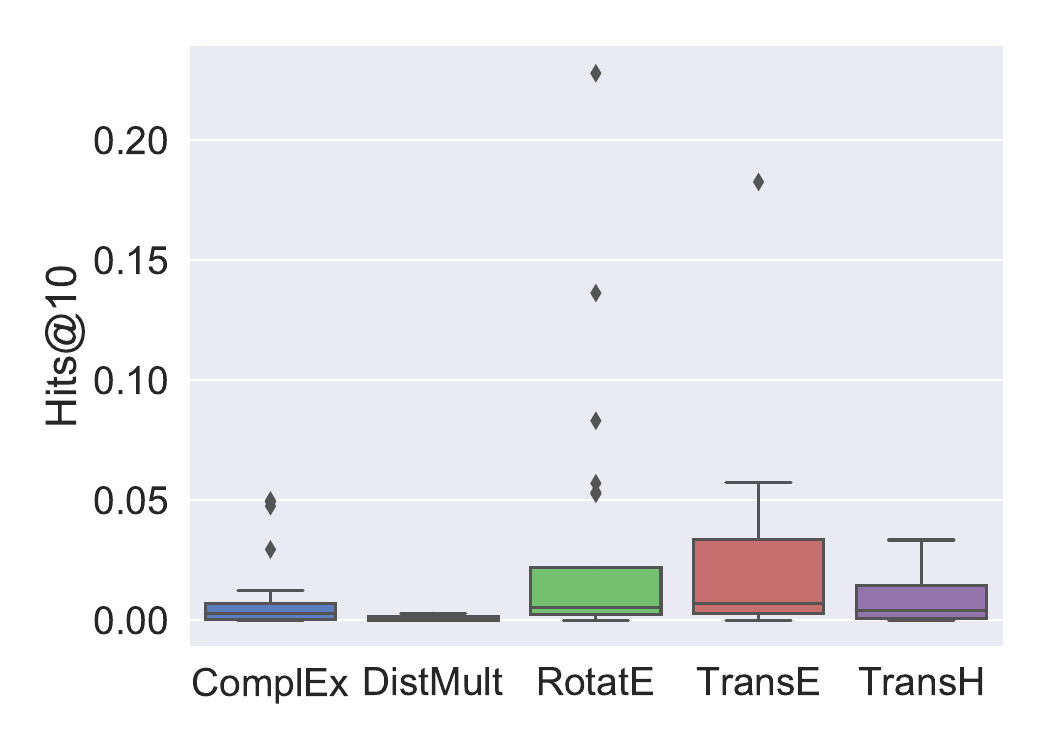}
        \caption{BioKG}\label{fig:ab:biokg}
    \end{subfigure}
    \caption{Distribution of the Hits@10 scores across all categorial training setup choices.}
    \label{fig:ab-model}
\end{figure}

Figure \ref{fig:ab-model} presents the distribution of Hits@10 scores on the test set across training setup choices for all models and both datasets, enabling a global view of how the models respond to changes in training setup. It can be seen that for Hetionet, many of the models have a similar range of performance, with DistMult standing out as having poor performance no matter the training setup, a trend which can also be seen on the BioKG dataset.

\begin{figure*}[!ht]
    \centering
    \begin{subfigure}[b]{0.45\textwidth}
        \centering
        \includegraphics[width=0.99\textwidth]{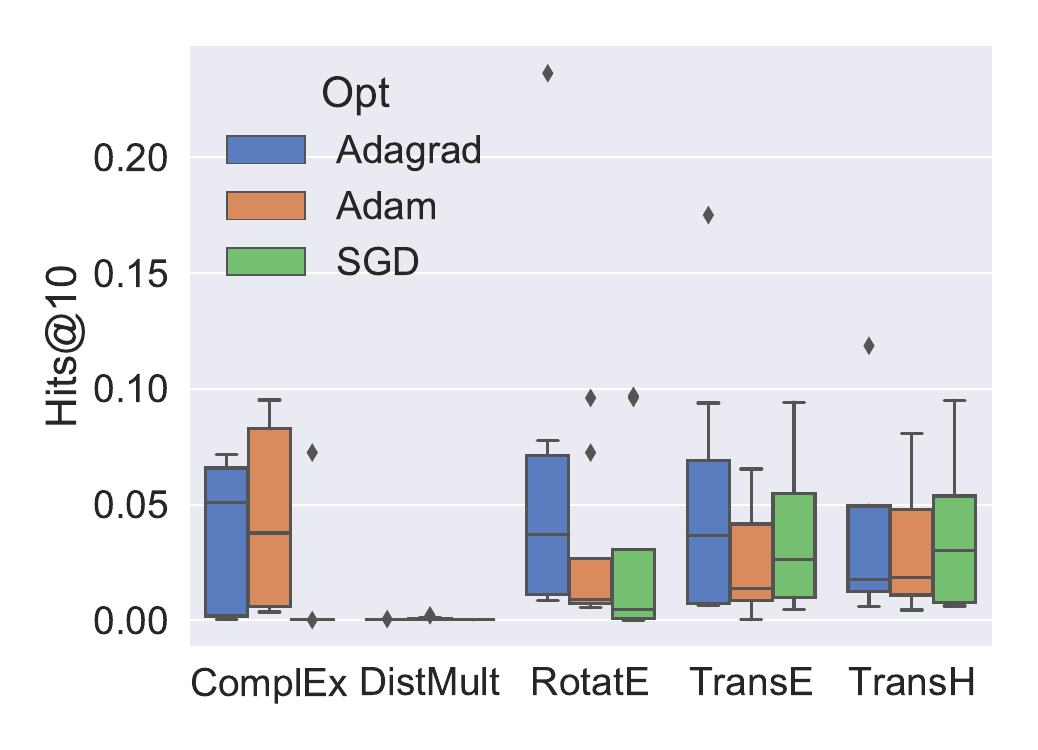}
        \caption{Hetionet Opt}\label{fig:ab:het:opt}
    \end{subfigure}
    \begin{subfigure}[b]{0.45\textwidth}
        \centering
        \includegraphics[width=0.99\textwidth]{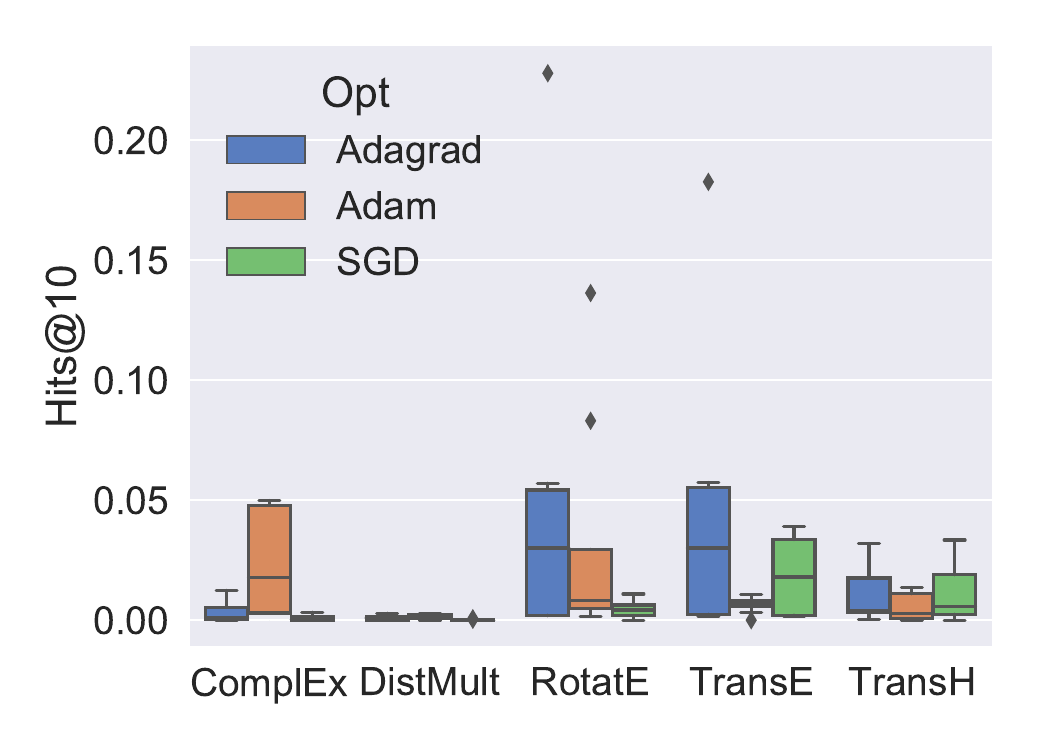}
        \caption{BioKG Opt}\label{fig:ab:biokg:opt}
    \end{subfigure}
    \begin{subfigure}[b]{0.45\textwidth}
        \centering
        \includegraphics[width=0.97\textwidth]{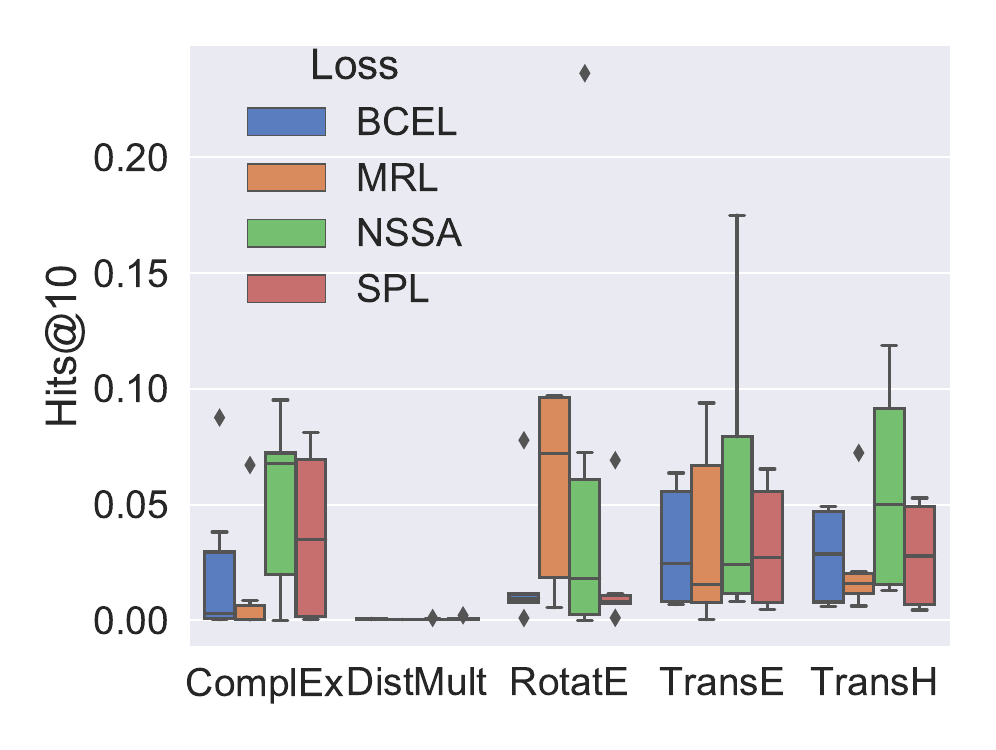}
        \caption{Hetionet Loss}\label{fig:ab:het:loss}
    \end{subfigure}
    \begin{subfigure}[b]{0.45\textwidth}
        \centering
        \includegraphics[width=0.97\textwidth]{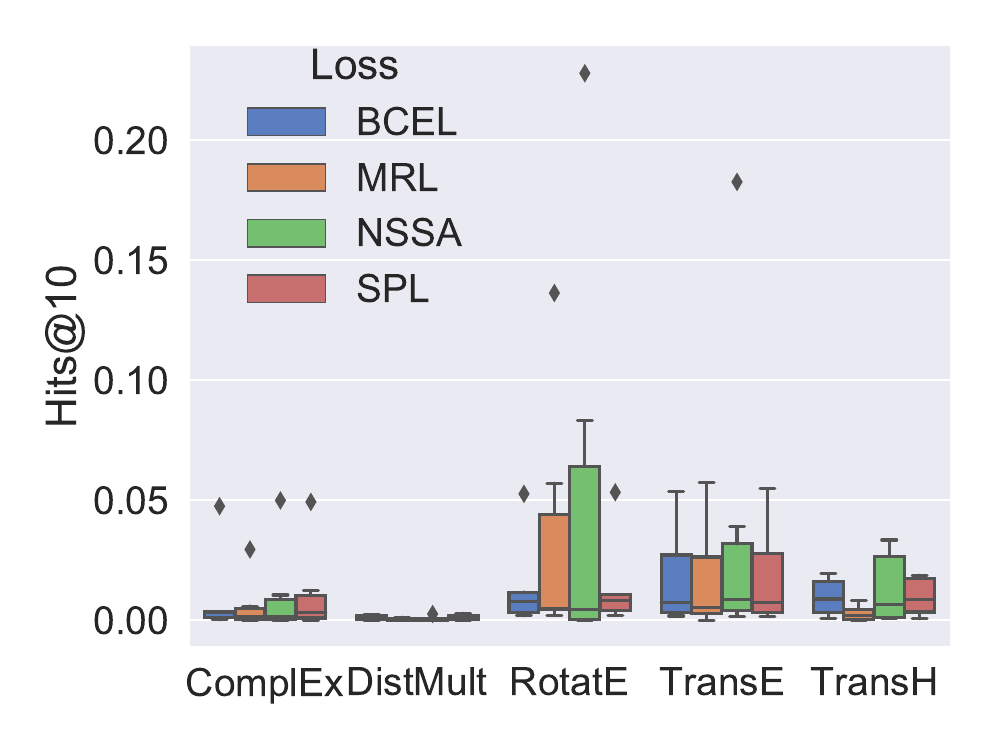}
        \caption{BioKG Loss}\label{fig:ab:biokg:loss}
    \end{subfigure}
    \begin{subfigure}[b]{0.45\textwidth}
        \centering
        \includegraphics[width=0.97\textwidth]{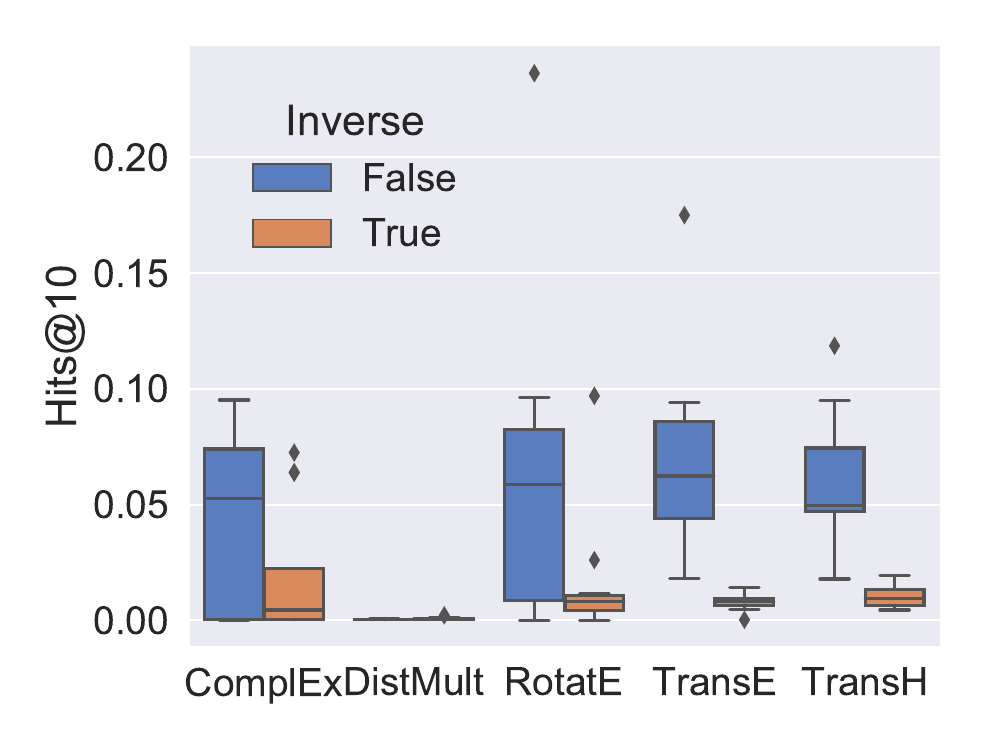}
        \caption{Hetionet Inverse}\label{fig:ab:het:inv}
    \end{subfigure}
    \begin{subfigure}[b]{0.45\textwidth}
        \centering
        \includegraphics[width=0.97\textwidth]{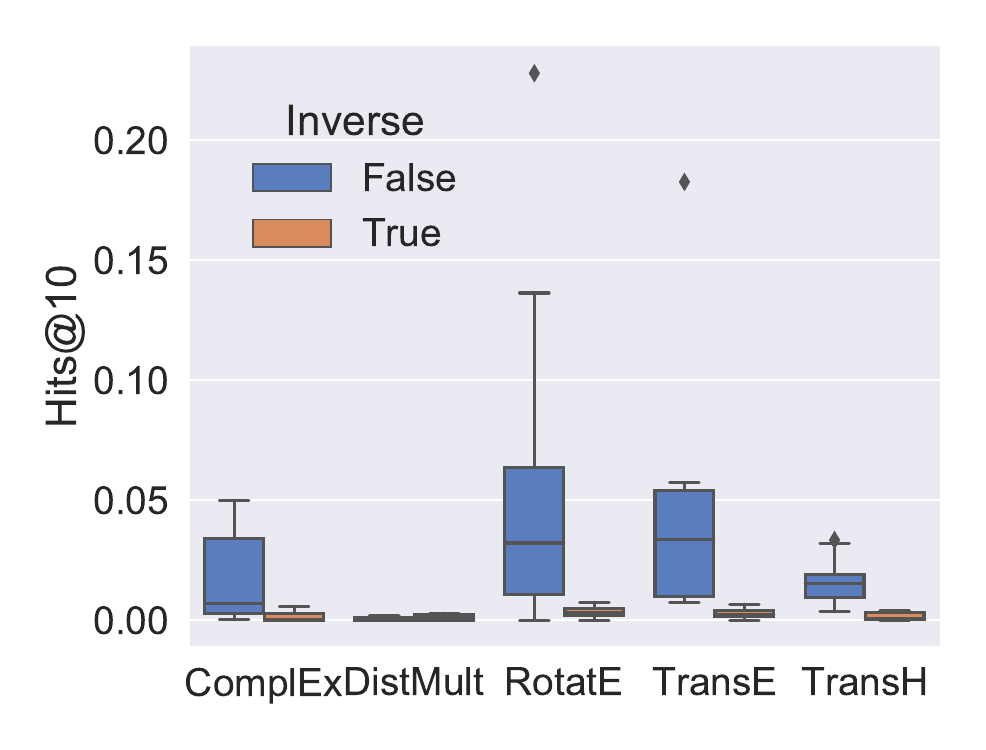}
        \caption{BioKG Inverse}\label{fig:ab:biokg:inv}
    \end{subfigure}
    \caption{The effect of different training setup choices across all models and both datasets.}\label{fig:ab-comp}
\end{figure*}

To further investigate the impact of different training setup choices, Figure \ref{fig:ab-comp} highlights the distributions of model performance separated by the the different choices for both datasets in which some interesting trends emerge. Perhaps the most striking observation can be seen in Figures \ref{fig:ab:het:inv} and \ref{fig:ab:biokg:inv}, which shows that adding inverse relationships to the training KG almost always performs worse on average than not including them -- this is in contrast to recent experimental evidence from non-biomedical domains~\cite{ali2020bringing}. Other observations include that, as shown in Figures \ref{fig:ab:het:loss} and \ref{fig:ab:biokg:loss}, using the NSSA loss function typically results in models having the highest peak predictive performance and that Figures \ref{fig:ab:het:opt} and \ref{fig:ab:biokg:opt} show that the Adagrad optimiser is typically used in the best performing training setup.

However, overall figure \ref{fig:ab-comp} reveals how multifactorial the problem of choosing the training setup can be -- clearly users must experiment to discover the most performant combination. The figure also highlights how improvements in the training setups have driven performance increases, perhaps more so than improvements in model architectures. For example, the RotatE model, given the correct training setup, is shown to perform best overall for both datasets, however it can also be outperformed by older approaches like TransE if suboptimal choices were made.

\subsection{Hyperparameter Optimisation}\label{ssec:hpo}

Even with the correct training setup, values given to key hyperparameters can have a significant impact on overall performance. It is common for various strategies to be employed to search for the best set of hyperparameters, a process called HyperParameter Optimisation (HPO)~\cite{akiba2019optuna}. In this section, we perform a detailed HPO search using two sampling strategies across all models and datasets. The two sampling strategies employed are the Bayesian \emph{Tree-structured Parzen Estimator Approach (TPE)}~\cite{bergstra2011algorithms}, an approach which creates a model to approximate the performance of a given hyperparameter set using historical information and a \emph{random} search~\cite{bergstra2012random} in which hyperparameters are sampled for each run independent of any previous ones. For each combination of search strategy, model and dataset we run 100 different experiments (with no time limit) to determine the hyperparameters - with the model seed, training setup (as detailed in Table~\ref{tab:default_parms}) and dataset split being fixed. The ranges of values searched over is detailed in Table~\ref{tab:hpo_ranges}, this range is the same for all models, datasets and sampling strategies. A given set of hyperparameters were assessed using the AMR metric for both search strategies on a holdout set of triples. This holdout set is a randomly selected, but fixed across all trails and repeats, set of triples comprising 10\% of all in the respective KGs.

\begin{table}[h!]
    \small
    \centering

    \begin{tabular}{l c }
        \toprule
        \textbf{Parameter} & \textbf{Value Range} \T\B  \\
        \midrule \midrule
        Embedding Dim      & \([16\ldots512,16]\)\T     \\
        Num Epochs         & \([100\ldots1000,100]\)    \\
        Learning Rate      & \([0.001\ldots0.1, \log]\) \\
        Num Negatives      & \([1\ldots100,10]\)        \\

        \bottomrule
    \end{tabular}
    \vspace{5pt}
    \caption{Range of search for parameter values, presented as min, max and step value.}\label{tab:hpo_ranges}
\end{table}

\begin{figure*}[!ht]
    \centering
    \begin{subfigure}[b]{0.48\textwidth}
        \centering
        \includegraphics[width=0.99\textwidth]{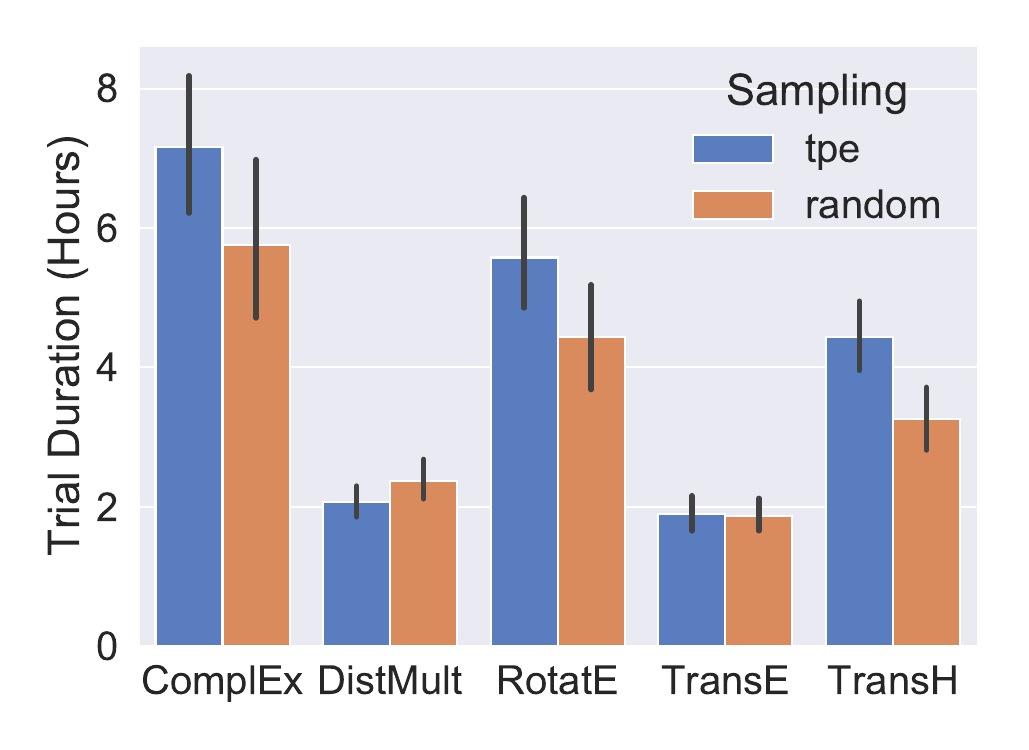}
        \caption{Trial Duration}\label{fig:trail-duration}
    \end{subfigure}
    \hfill
    \begin{subfigure}[b]{0.48\textwidth}
        \centering
        \includegraphics[width=0.99\textwidth]{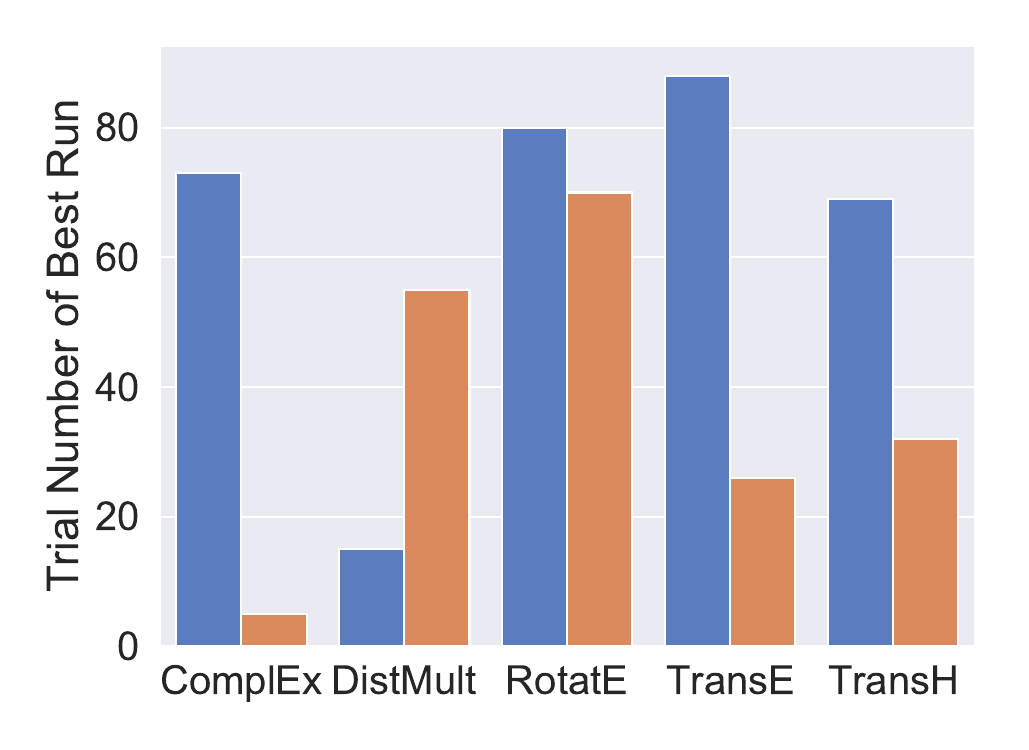}
        \caption{Best Trial Number}\label{fig:trail-best}
    \end{subfigure}
    \hfill
    \begin{subfigure}[b]{0.48\textwidth}
        \centering
        \includegraphics[width=0.99\textwidth]{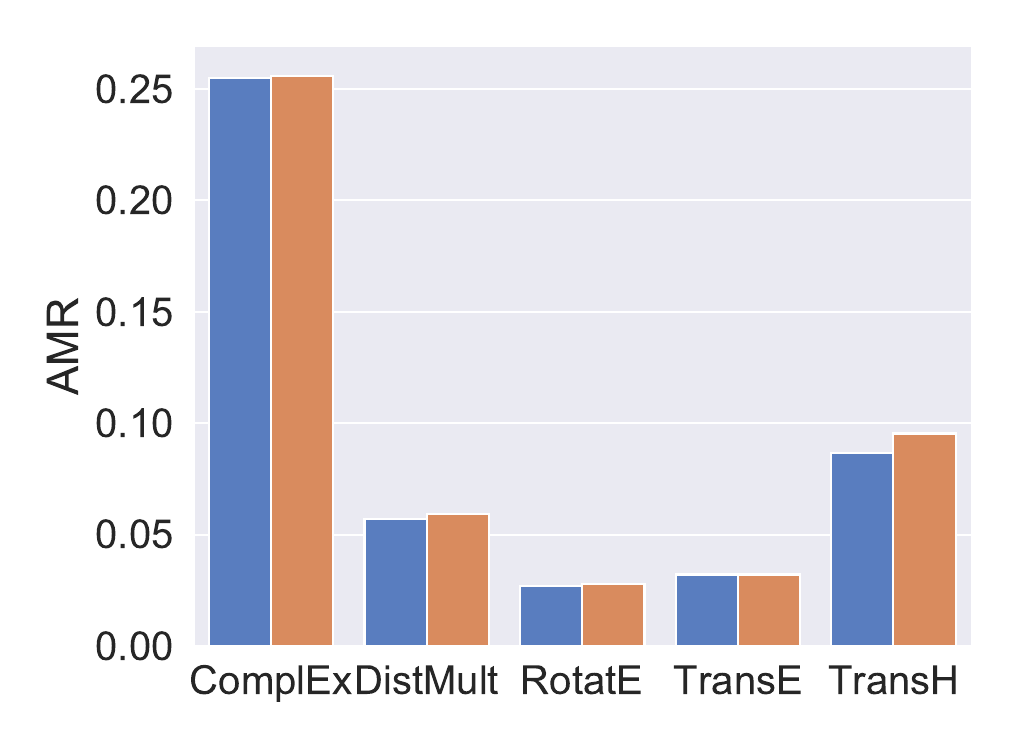}
        \caption{Best Trial (AMR)}\label{fig:trial-amr}
    \end{subfigure}
    \hfill
    \begin{subfigure}[b]{0.48\textwidth}
        \centering
        \includegraphics[width=0.99\textwidth]{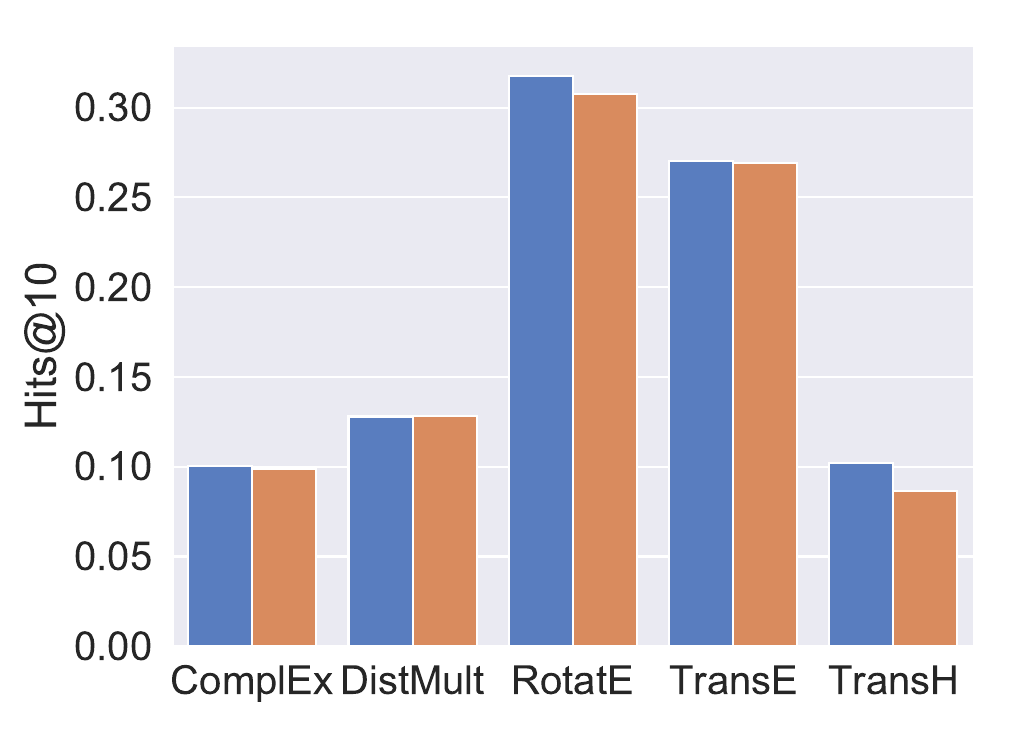}
        \caption{Best Trial (Hits@10)}\label{fig:trial-h10}
    \end{subfigure}

    \caption{Performance of the two sampling algorithms after 100 trials across a range of metrics with all models on Hetionet.}\label{fig:hpo-hetnet}
\end{figure*}

Figure \ref{fig:hpo-hetnet} presents an overview of the HPO experiments comparing the two sampling strategies, focusing on the Hetionet dataset for brevity (the results for BioKG demonstrated very similar patterns). Firstly, Figure \ref{fig:trail-duration} displays the mean runtime of all models for both sampling strategies, showing that the TPE approach generated longer trial durations on average. This is likely as it tuned parameters which can increase runtime, embedding dimension, number of negative samples and number of epochs, in the pursuit of additional predictive performance. Figure~\ref{fig:trail-best} shows on which of the 100 experimental trials the best performing model was produced, with the TPE approach generally displaying its best performance close to the maximum number of repeats. Figures~\ref{fig:trial-amr} and \ref{fig:trial-h10} display the predictive performance of the approaches as measured by AMR and Hits@10 respectively. One surprising observation is how close the performance is of the TPE and random sampling strategies across all models -- with TPE only producing models which are marginally better than those trained on random hyperparameter choices. This highlights how given enough repeats, a random search can happen upon near optimal parameters by chance, at least for these models and datasets. Indeed given the additional average runtime incurred by TPE, a random search may be the better balance of runtime and performance. One final observation is the negative correlation between the AMR and Hits@10 performance, which is encouraging to see as the HPO search was only optimising the AMR value.

\begin{figure}[!ht]
    \centering
    \begin{subfigure}[b]{0.48\textwidth}
        \centering
        \includegraphics[width=0.99\textwidth]{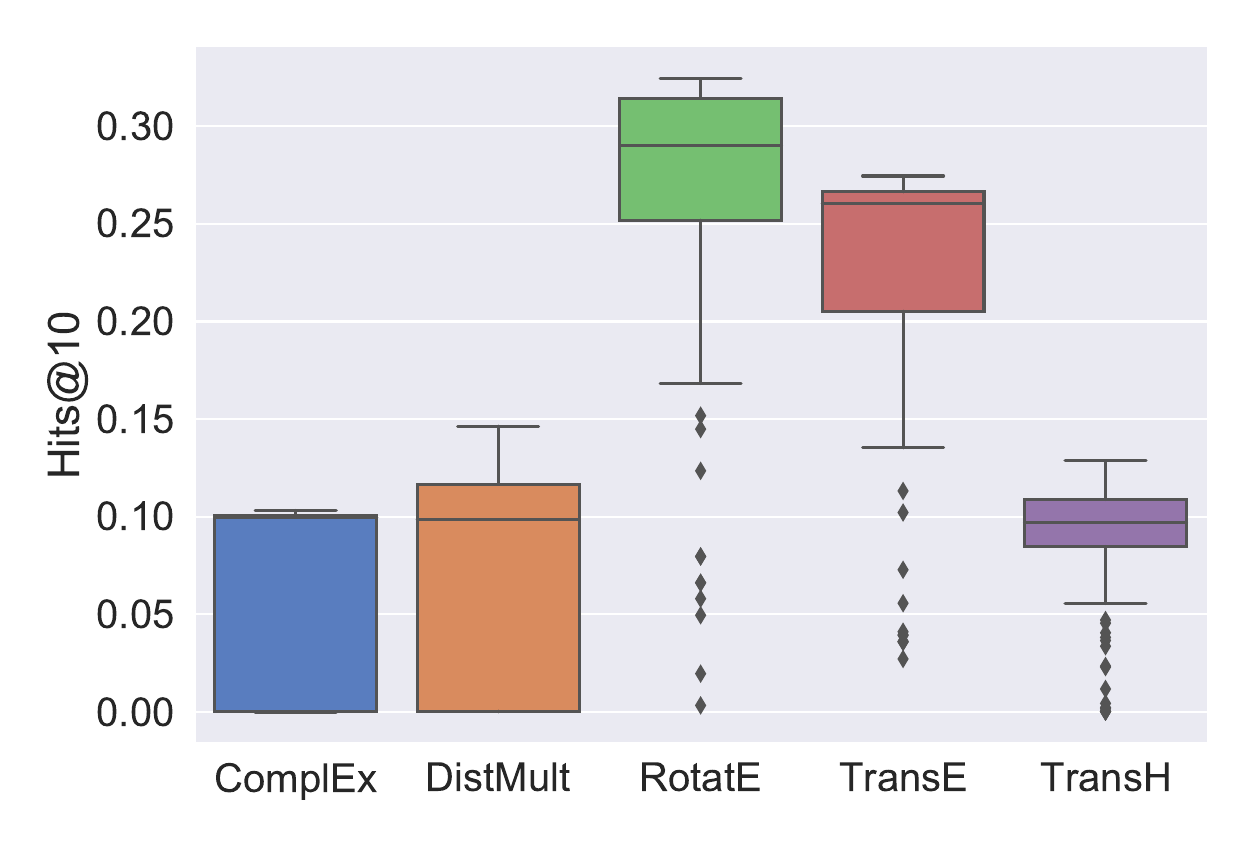}
        \caption{Hetionet}\label{fig:hpo:dist:hetnet}
    \end{subfigure}
    \begin{subfigure}[b]{0.48\textwidth}
        \centering
        \includegraphics[width=0.98\textwidth]{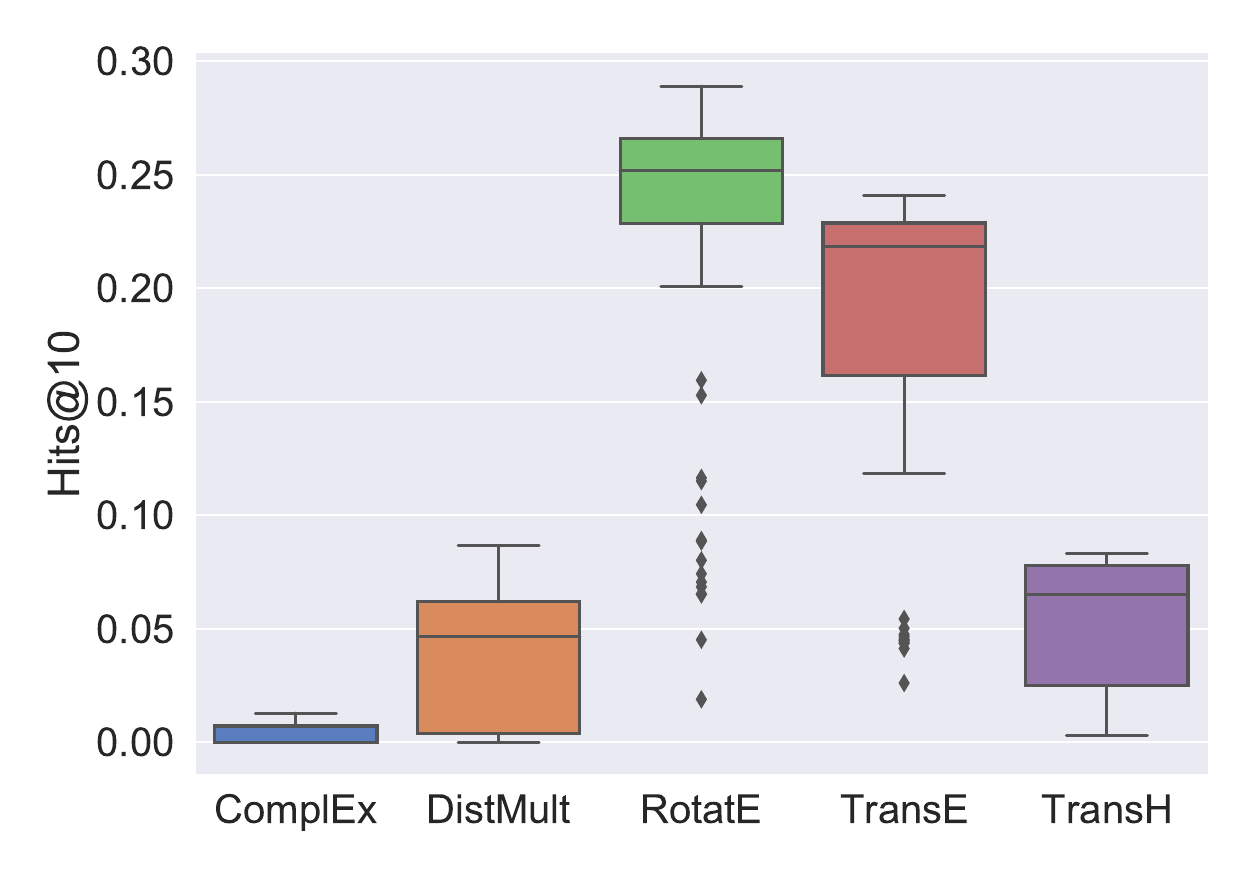}
        \caption{BioKG}\label{fig:hpo:dist:biokg}
    \end{subfigure}

    \caption{Distribution of the Hits@10 scores across all 100 runs of different parameters.}\label{fig:hpo-dist}

\end{figure}

In Figure~\ref{fig:hpo-dist}, the distribution of Hits@10 scores across all models and 100 trials using TPE sampling is presented. The figure shows the importance of correct hyperparameter selection, as all models are shown to be sensitive to them. Indeed the selection of hyperparameters could change the ordering of ranked model performance, with the best performing model RotatE demonstrating performance  below others given suboptimal parameter choices. Additionally one should consider that older approaches like TransE can still perform very competitively given appropriate time is spent tuning them. This is important to consider as new models continue to be proposed, and we hope that authors make the effort to appropriately tune baseline approaches for fairer comparison.

\begin{figure*}[!ht]
    \centering
    \begin{subfigure}[b]{0.8\textwidth}
        \centering
        \includegraphics[width=0.99\textwidth]{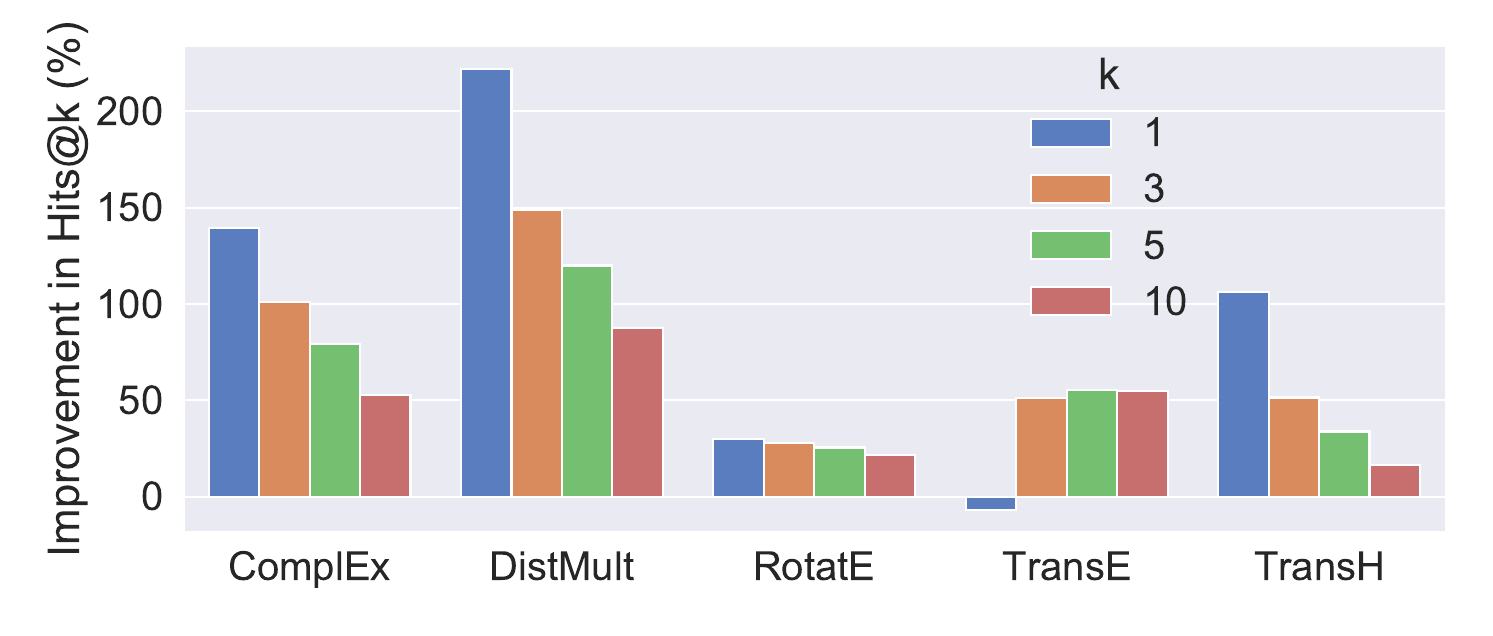}
        \caption{Hetionet}\label{fig:hpo:imp:hetnet}
    \end{subfigure}
    \begin{subfigure}[b]{0.8\textwidth}
        \centering
        \includegraphics[width=0.99\textwidth]{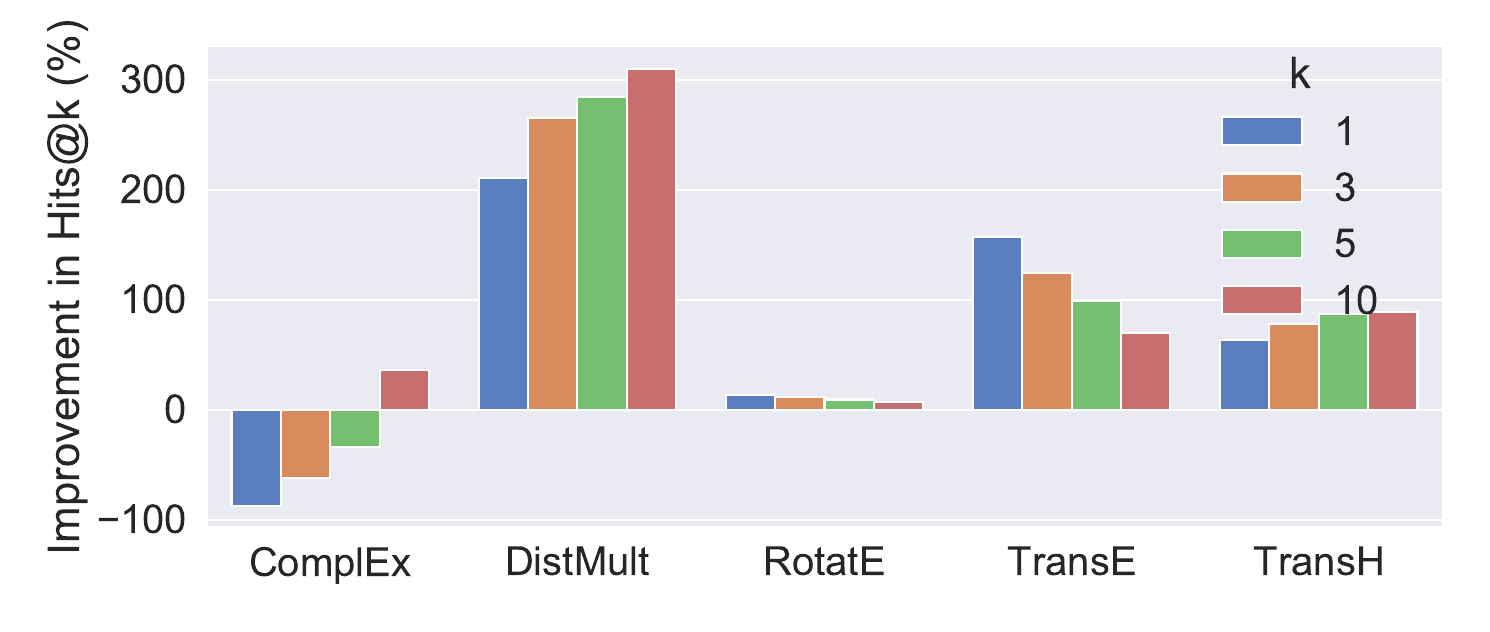}
        \caption{BioKG}\label{fig:hpo:imp:biokg}
    \end{subfigure}

    \caption{Percentage improvement for the best parameter configurations over baseline values.}\label{fig:hpo-list}

\end{figure*}

Figure~\ref{fig:hpo-list} highlights the percentage in improvement across different Hits@k levels for the best set of parameters from the TPE sampling strategy over the baseline ones (i.e., models trained using the setup defined in Table \ref{tab:default_parms}). It can be seen that the majority of models clearly benefit from a detailed hyperparameter search, albeit to varying degrees. For example, those approaches which have worst performance overall such as DistMult benefit most by having their parameters tuned, conversely the higher performing approaches like RotatE benefit to a lesser degree. One observation seen in Figure~\ref{fig:hpo:imp:biokg} is that the best set of parameters produced via the HPO process fails to exceed the performance of the baseline at lower values of k for the ComplEx model. However, overall these experiments demonstrate that time should be taken to fine-tune hyperparameters as the performance increase over even sensible defaults can be dramatic.

\subsubsection{Hyperparameter Values}

\begin{figure}[!ht]
    \centering
    \includegraphics[width=0.65\textwidth]{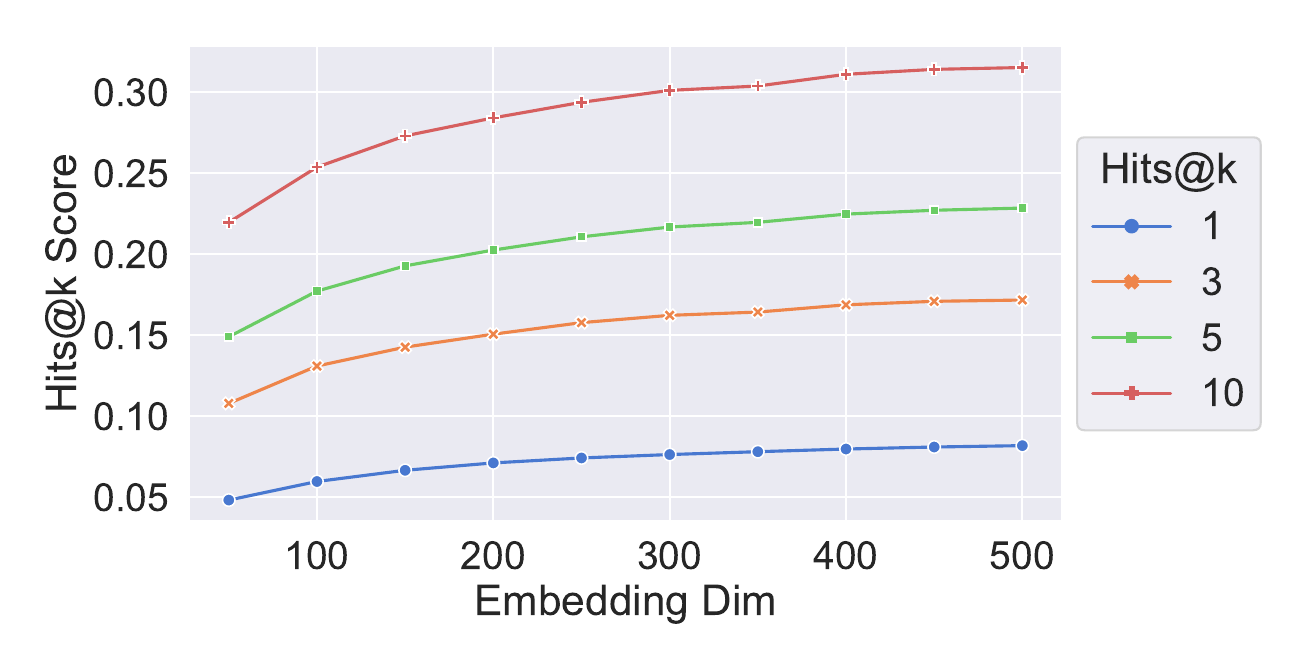}
    \caption{Performance changes with embedding dimension for RotatE on Hetionet, whilst fixing all other parameters.}
    \label{fig:emb_biokg}
\end{figure}

\begin{figure*}[!ht]
    \centering
    \begin{subfigure}[b]{0.45\textwidth}
        \centering
        \includegraphics[width=0.99\textwidth]{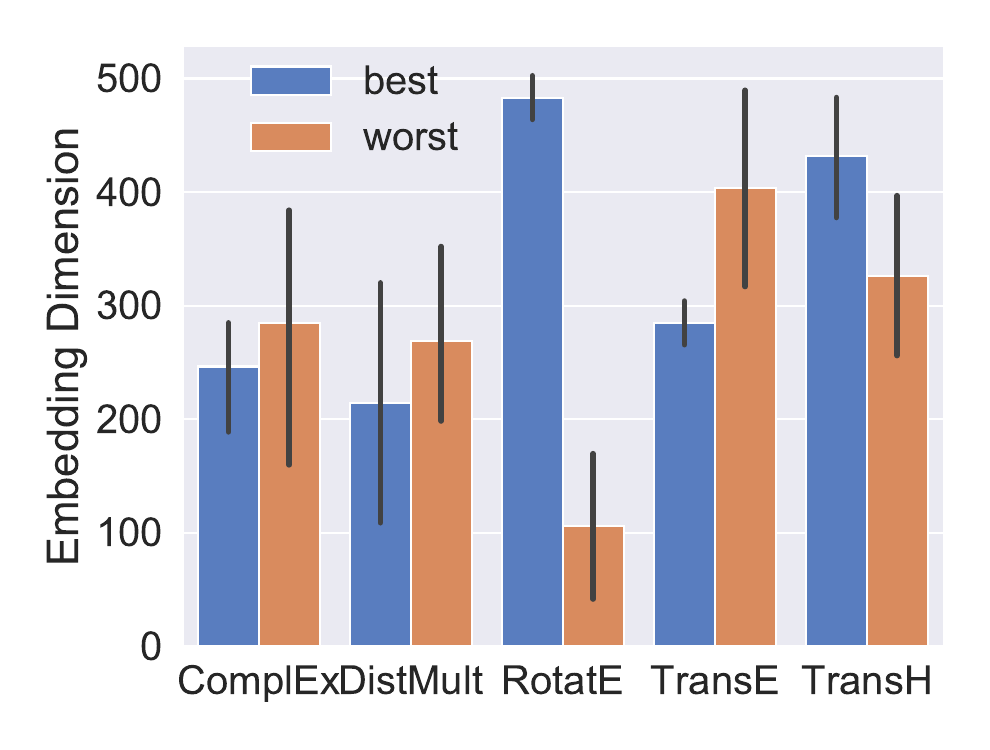}
        \caption{Embedding Size}\label{fig:hpo_emb}
    \end{subfigure}
    \hfill
    \begin{subfigure}[b]{0.45\textwidth}
        \centering
        \includegraphics[width=0.99\textwidth]{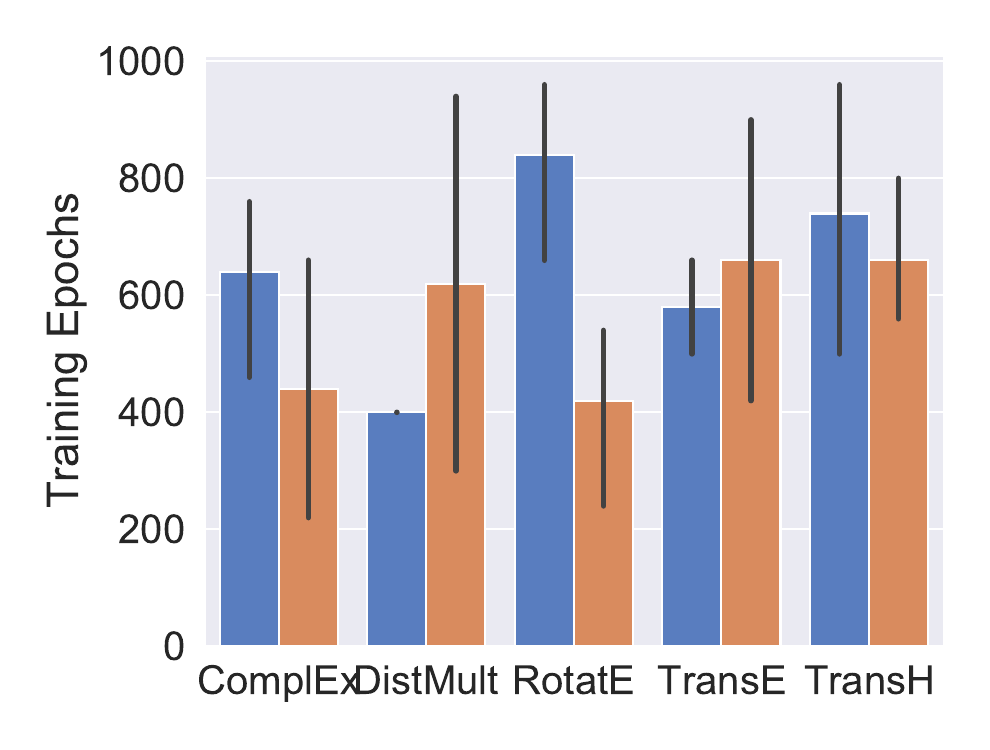}
        \caption{Training Epochs}\label{fig:hpo_epoch}
    \end{subfigure}
    \hfill
    \begin{subfigure}[b]{0.45\textwidth}
        \centering
        \includegraphics[width=0.99\textwidth]{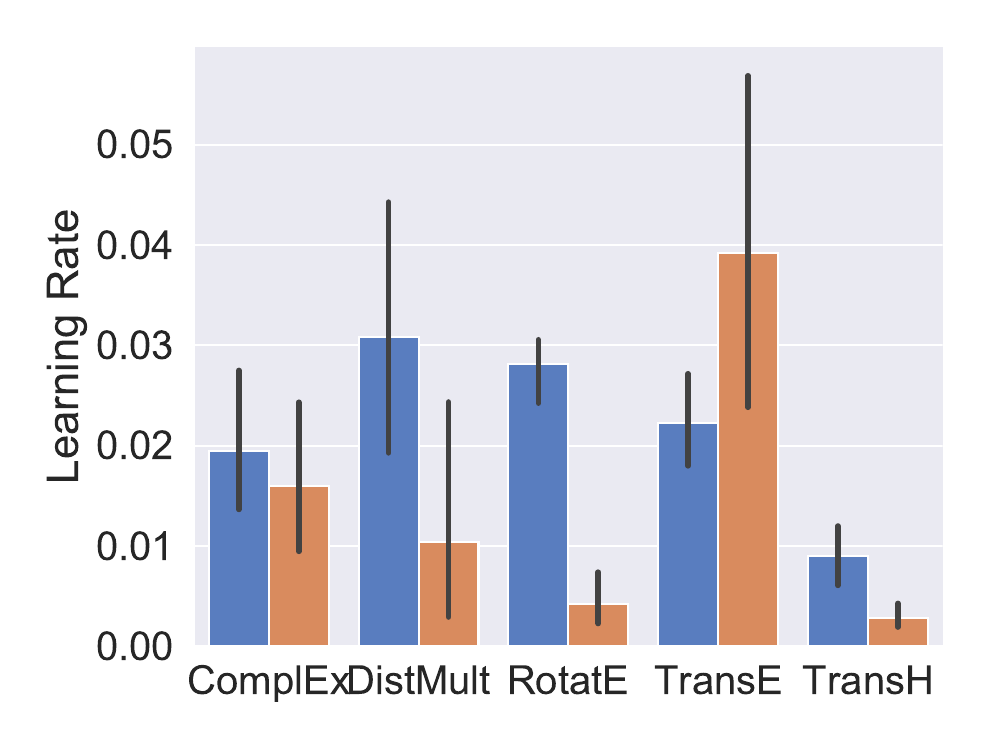}
        \caption{Learning Rate}\label{fig:hpo_lr}
    \end{subfigure}
    \hfill
    \begin{subfigure}[b]{0.45\textwidth}
        \centering
        \includegraphics[width=0.99\textwidth]{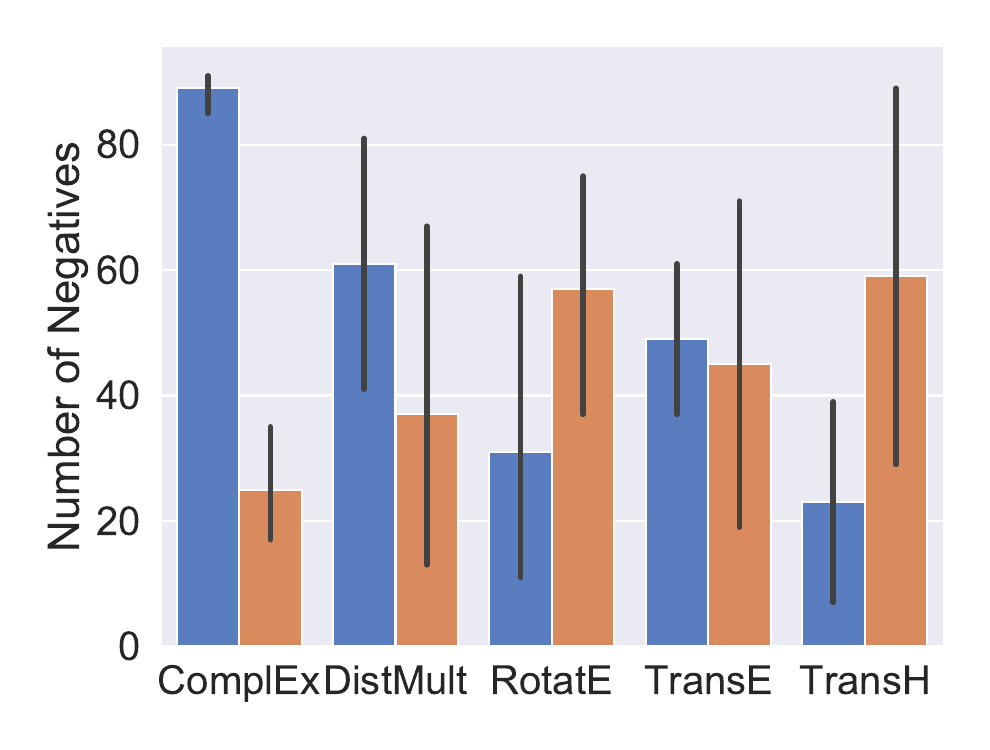}
        \caption{Num Negatives}\label{fig:hpo_neg}
    \end{subfigure}

    \caption{The mean hyperparameter values of the top five best and worst configurations by model on the Hetionet dataset.}\label{fig:hpo-parms}

\end{figure*}

To investigate the values taken by hyperparameters and how they correlate with performance, we averaged hyperparameter values for the five best and worst configurations (as determined by the Hits@10 metric) from the HPO runs for both search approaches on the Hetionet dataset -- which can be seen in Figure \ref{fig:hpo-parms}. Figure~\ref{fig:hpo-parms} reveals some interesting trends, for example it can seen that there is a large variance in many hyperparameter values even among just five examples, indicating that there are multiple permutations of hyperparameter values which perform to a similar level. It is also surprising how close the range of values for many parameters are for the five best and worst configurations, perhaps revealing how multifactorial the problem of choosing hyperparameters to be -- performance seems to be driven by specific combinations of values given to multiple parameters. This indicates that researchers should search over multiple parameters simultaneously, rather than focusing on a single one to achieve the best overall configuration (Although Figure \ref{fig:emb_biokg} shows that model performance can be increased via changes in a single parameter, although it does not explain all). However, there are examples in the figure of clear differences between best and worst configurations, with RotatE performing better on average with larger embeddings, more training epochs and a larger learning rate.

The average parameter values over the five best runs for both dataset is displayed in Table~\ref{tab:hpo_best}. This shows that, despite the datasets being from the same domain, parameter values are not similar, highlighting that values need to be optimised on a per-dataset basis. Comparing the values for the two KGs some trends do emerge however, for example all models on BioKG tend to require a larger learning rate and have a broader range of embedding dimensions which perform well.

\begin{table}[h!]
    \tiny
    \centering
    \begin{tabular}{l l c c c c c }
        \toprule
        \textbf{Dataset}          & \textbf{Approach} & \multicolumn{5}{c}{\textbf{Top Five Configs}}\T\B                                                                            \\
        \midrule \midrule
                                  &                   & Emb Size                                          & Num Epochs    & Learning Rate     & Num Neg     & Hits@10 \(\uparrow\)\B \\
        \cline{3-7}

        \multirow{5}{*}{Hetionet} & ComplEx           & 246\(\pm\)61                                      & 640\(\pm\)195 & 0.019\(\pm\)0.009 & 89\(\pm\)4  & 0.102\(\pm\)0.001\T    \\
                                  & DistMult          & 214\(\pm\)141                                     & 400\(\pm\)000 & 0.030\(\pm\)0.015 & 61\(\pm\)27 & 0.140\(\pm\)0.004      \\
                                  & RotatE            & 483\(\pm\)26                                      & 840\(\pm\)194 & 0.028\(\pm\)0.004 & 31\(\pm\)30 & 0.322\(\pm\)0.001      \\
                                  & TransE            & 285\(\pm\)26                                      & 580\(\pm\)110 & 0.022\(\pm\)0.006 & 49\(\pm\)16 & 0.272\(\pm\)0.001      \\
                                  & TransH            & 432\(\pm\)67                                      & 740\(\pm\)297 & 0.008\(\pm\)0.004 & 23\(\pm\)20 & 0.129\(\pm\)0.076\B    \\
        \midrule
        \multirow{5}{*}{BioKG}    & ComplEx           & 422\(\pm\)96                                      & 600\(\pm\)187 & 0.058\(\pm\)0.030 & 81\(\pm\)10 & 0.012\(\pm\)0.001\T    \\
                                  & DistMult          & 349\(\pm\)180                                     & 100\(\pm\)000 & 0.039\(\pm\)0.006 & 71\(\pm\)0  & 0.082\(\pm\)0.002      \\
                                  & RotatE            & 413\(\pm\)51                                      & 760\(\pm\)89  & 0.055\(\pm\)0.020 & 25\(\pm\)9  & 0.286\(\pm\)0.002      \\
                                  & TransE            & 378\(\pm\)115                                     & 540\(\pm\)134 & 0.078\(\pm\)0.019 & 91\(\pm\)0  & 0.239\(\pm\)0.001      \\
                                  & TransH            & 317\(\pm\)114                                     & 920\(\pm\)130 & 0.067\(\pm\)0.018 & 27\(\pm\)9  & 0.08\(\pm\)0.000\B     \\

        \bottomrule
    \end{tabular}
    \vspace{5pt}
    \caption{The top five best performing hyperparameter configurations (includes models from both TPE and random search) for all models and datasets. Values presented as mean and standard deviation.}\label{tab:hpo_best}

\end{table}

In Table~\ref{tab:hpo_best_overall} we report the specific set of hyperparameters that resulted in the best overall performing models across both Hetionet and BioKG. These are compared with hyperparameters, taken from the original publications, for two non-biomedical KGs: FB15K~\cite{bordes2013translating} and WN18~\cite{bordes2013translating}. The table highlights how the parameters across the two domains are rarely comparable in value. For example, typically all models require a larger embedding size and more negative samples to achieve the highest level of performance using biomedical KGs. This further reinforces the idea that one cannot assume hyperparameters will generalise across domains, or even across datasets from within a domain. This can be seen in the difference in parameters for the same model across Hetionet and BioKG, despite the two comprising similar entity and relation types.

\begin{table}[h!]
    \tiny
    \centering
    \begin{tabular}{l l c c c c }
        \toprule
        \textbf{Approach}         & \textbf{Dataset} & \multicolumn{4}{c}{\textbf{Best Config}}\T\B                                          \\
        \midrule \midrule
                                  &                  & Emb Size                                     & Num Epochs & Learning Rate & Num Neg\B \\
        \cline{3-6}

        \multirow{4}{*}{ComplEx}  & Hetionet         & 272                                          & 700        & 0.03          & 91\T      \\
                                  & BioKG            & 464                                          & 600        & 0.09          & 91\B      \\
        \cline{2-6}
                                  & FB15K            & 200                                          & 1000       & 0.5           & 10\T      \\
                                  & WN18             & 150                                          & 1000       & 0.5           & 1\B       \\
        \midrule
        \multirow{4}{*}{DistMult} & Hetionet         & 80                                           & 400        & 0.02          & 41 \T     \\
                                  & BioKG            & 480                                          & 100        & 0.05          & 71\B      \\
        \cline{2-6}
                                  & FB15K            & 100                                          & 100        & 0.1           & 2\T       \\
                                  & WN18             & 100                                          & 300        & 0.1           & 2\B       \\
        \midrule
        \multirow{4}{*}{RotatE}   & Hetionet         & 512                                          & 500        & 0.03          & 41 \T     \\
                                  & BioKG            & 448                                          & 900        & 0.06          & 31\B      \\
        \cline{2-6}
                                  & FB15K            & 1000                                         & 1000       & 0.0001        & 128\T     \\
                                  & WN18             & 500                                          & 1000       & 0.0001        & 1024\B    \\
        \midrule
        \multirow{4}{*}{TransE}   & Hetionet         & 304                                          & 500        & 0.02          & 61 \T     \\
                                  & BioKG            & 448                                          & 600        & 0.1           & 91\B      \\
        \cline{2-6}
                                  & FB15K            & 50                                           & 1000       & 0.01          & 1\T       \\
                                  & WN18             & 20                                           & 4000       & 0.01          & 1\B       \\
        \midrule
        \multirow{4}{*}{TransH}   & Hetionet         & 480                                          & 800        & 0.005         & 1 \T      \\
                                  & BioKG            & 368                                          & 900        & 0.06          & 31\B      \\
        \cline{2-6}
                                  & FB15K            & 100                                          & 500        & 0.005         & 1\T       \\
                                  & WN18             & 50                                           & 500        & 0.01          & 1\B       \\
        \bottomrule
    \end{tabular}
    \vspace{5pt}
    \caption{The hyperparameters from the best performing run. Results from two non-biomedical KGs (FB15k and WN18) are included for comparison purposes.}\label{tab:hpo_best_overall}
\end{table}

\subsubsection{Task Specific HPO}\label{sssection:task-specific-hpo}

In many applications, achieving good performance on specific relation types is the ultimate goal, with the rest of the graph hopefully being used to improve this specific task. One interesting aspect to consider is if tuning the HPO search to focus on just the relations of interest generates better performing models than tuning on all possible relation types. For example, if we were specifically interested in being able to predict gene to disease edges, would tuning the HPO process on just these edges yield better overall performance on this task? To investigate this, we repeated the HPO search using the TPE method on Hetionet across both RotatE (generally the best performing model thus far) and ComplEx (typically one of the lower performing models), where parameters were tuned via model performance only on gene to disease edges. This is in contrast to the previous HPO search, where parameters were tuned via model performance measured on a random holdout set comprised, potentially, of all relation types.

\begin{figure}[!ht]
    \centering
    \begin{subfigure}[b]{0.43\textwidth}
        \centering
        \includegraphics[width=0.99\textwidth]{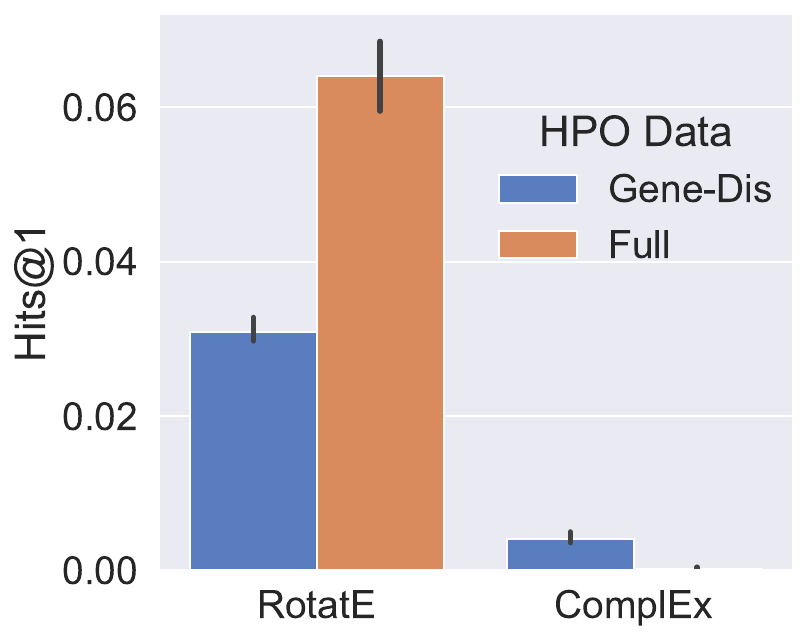}
        \caption{Hits@1}\label{fig:gd-hpo:1}
    \end{subfigure}
    \hfill
    \begin{subfigure}[b]{0.43\textwidth}
        \centering
        \includegraphics[width=0.99\textwidth]{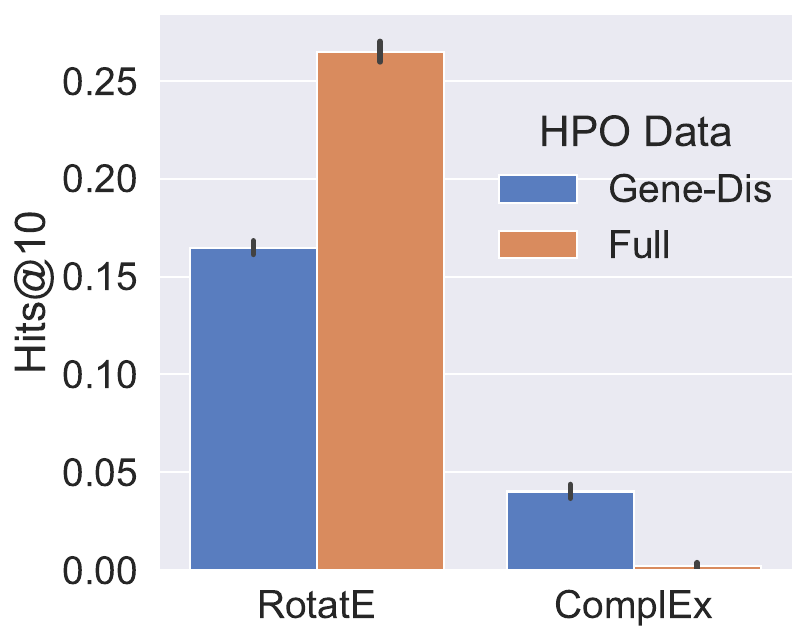}
        \caption{Hits@10}\label{fig:gd-hpo:10}
    \end{subfigure}

    \caption{Comparison between models trained on Hetionet using HPO values taken from optimising with TPE across all relation types (Full) and just Gene-Disease relations (Gene-Dis). Hits@1 and Hits@10 scores are presented on a holdout testset comprised only of Gene-Disease relations.}\label{fig:gd-hpo}
\end{figure}

This comparison is presented in Figure \ref{fig:gd-hpo}, where RotatE and ComplEx were evaluated using the Hits@1 and Hits@10 metrics on a random selection of gene-disease edges using hyperparameters optimised for this edge type, compared with those optimised on all edges. The figure shows that for ComplEx, which performs poorly on gene-disease edges when using the hyperparameters tuned on the full set of edges, its predictive performance increases when using the edge-tuned hyperparameters. The gene-disease edges represent a small fraction of total in Hetionet at less than 1\%, which seems to impact the two models in different ways. Compared to ComplEx, RotatE demonstrates a decrease in performance using the gene-disease tuned hyperparameters, suggesting it is seemingly not being able to discover an optimal set of hyperparameters using the much sparser signal offered by the reduced edge counts. It is also possible that, given the much smaller size of the validation set used for hyperparameter tuning, that the model was over-fitting to these limited edges and not being able to generalise to unseen examples. Overall these experiments suggest that, when a model performs well overall, hyperparameters learned across all relation types can still be optimal for use in a relation type specific setting.

\subsection{Model Initialisation Random Seed}

We now assess how model predictive performance changes as the random seed used to initialise model parameters is varied. Many architectures are known to be impacted by the initial conditions of the parameters as determined by the random seed, with performance not being consistent over seeds~\cite{madhyastha2019model}. However, thus far, the impact of model random seed has not been thoroughly investigated in the biomedical KG domain, with results in the literature rarely being presented as averaged over different random seeds.

To assess the impact of model initialisation, in this experiment, we keep all other experiment variables constant (fixing the hyperparameters, training setup and dataset split) and run repeats over a fixed set of 20 random seed for all models and across both datasets. The overall results of this are presented in Table~\ref{tab:model-seed}, which shows that the majority of the models are relatively robust to the random seed used for parameter initialisation. The one clear exception is the performance of the DistMult model, which demonstrates a large sensitivity to the random seed with certain metrics.

\begin{table}[h!]
    \tiny
    \centering

    \begin{tabular}{l l  c c c  c c c  }
        \toprule
        \textbf{Dataset}          & \textbf{Approach} & \multicolumn{4}{c}{\textbf{Metric}} \T\B                                                                                        \\
        \midrule \midrule
                                  &                   & AMR \(\downarrow\)                       & MRR \(\uparrow\)           & Hits@1 \(\uparrow\)        & Hits@10 \(\uparrow\)\B     \\
        \cline{3-6}

        \multirow{5}{*}{Hetionet} & ComplEx           & 0.167\(\pm\)0.009                        & 0.026\(\pm\)0.009          & 0.008\(\pm\)0.003          & 0.064\(\pm\)0.024\T        \\
                                  & DistMult          & 0.201\(\pm\)0.303                        & 0.036\(\pm\)0.019          & 0.012\(\pm\)0.007          & 0.079\(\pm\)0.045          \\
                                  & RotatE            & \textbf{0.035\(\pm\)0.000}               & \textbf{0.127\(\pm\)0.000} & \textbf{0.063\(\pm\)0.000} & \textbf{0.262\(\pm\)0.001} \\
                                  & TransE            & 0.053\(\pm\)0.000                        & 0.079\(\pm\)0.001          & 0.034\(\pm\)0.001          & 0.117\(\pm\)0.002          \\
                                  & TransH            & 0.126\(\pm\)0.000                        & 0.033\(\pm\)0.001          & 0.007\(\pm\)0.000          & 0.088\(\pm\)0.002\B        \\
        \midrule
        \multirow{5}{*}{BioKG}    & ComplEx           & 0.213\(\pm\)0.011                        & 0.008\(\pm\)0.001          & 0.003\(\pm\)0.000          & 0.008\(\pm\)0.003\T        \\
                                  & DistMult          & 0.560\(\pm\)0.339                        & 0.015\(\pm\)0.003          & 0.007\(\pm\)0.001          & 0.027\(\pm\)0.006          \\
                                  & RotatE            & 0.022\(\pm\)0.000                        & \textbf{0.123\(\pm\)0.000} & \textbf{0.059\(\pm\)0.000} & \textbf{0.240\(\pm\)0.001} \\
                                  & TransE            & \textbf{0.021\(\pm\)0.000}               & 0.062\(\pm\)0.000          & 0.019\(\pm\)0.000          & 0.134\(\pm\)0.001          \\
                                  & TransH            & 0.078\(\pm\)0.001                        & 0.022\(\pm\)0.000          & 0.008\(\pm\)0.000          & 0.042\(\pm\)0.001\B        \\

        \bottomrule
    \end{tabular}
    \vspace{5pt}
    \caption{The mean performance with standard deviation over 10 fixed random seeds of all models and both datasets as measured by various metrics. Here only the random seed used to initialise the model parameters is changed, whilst all over variables are kept fixed.}\label{tab:model-seed}

\end{table}

\begin{figure}[!ht]
    \centering
    \begin{subfigure}[b]{0.45\textwidth}
        \centering
        \includegraphics[width=0.99\textwidth]{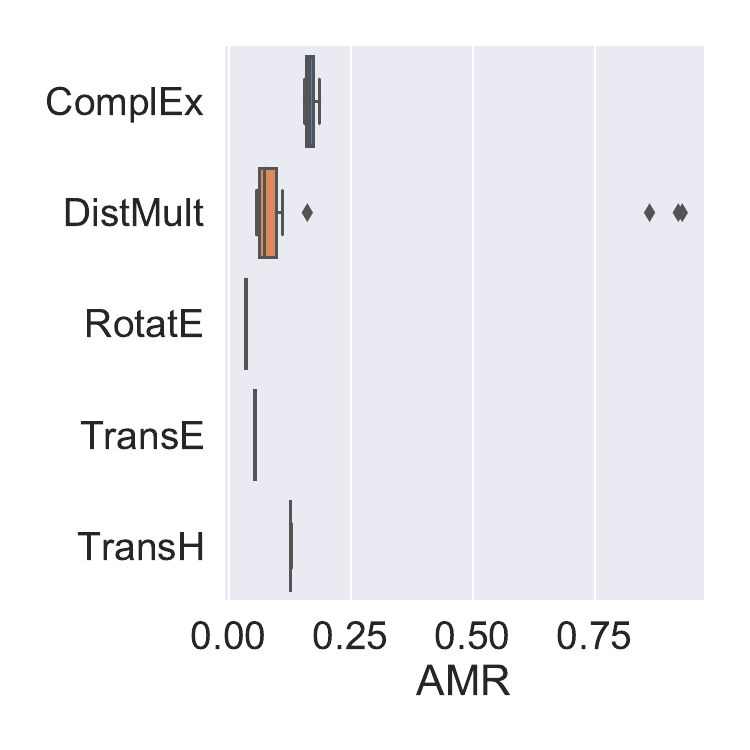}
        \caption{Hetionet}\label{fig:ms:hetnet}
    \end{subfigure}
    \begin{subfigure}[b]{0.45\textwidth}
        \centering
        \includegraphics[width=0.99\textwidth]{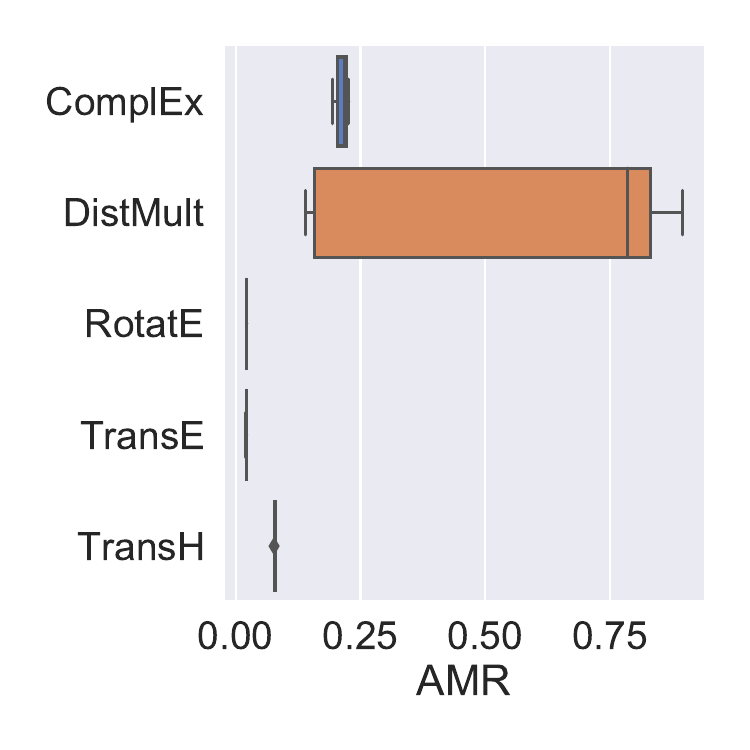}
        \caption{BioKG}\label{fig:ms:biokg}
    \end{subfigure}

    \caption{Distribution of AMR score obtained by repeating the experiment whilst varying the model random seed.}\label{fig:model-seed}

\end{figure}

To investigate this further, the distribution of the AMR score across the random seeds is presented in Figure~\ref{fig:model-seed}, showing that DistMult's average performance is skewed by a few outlying poor results at certain seed values. Although it can be seen that, whilst ignoring the outliers for Hetionet brings DistMult's performance more inline with the other models, the majority of the runs for the BioKG dataset are still significantly worse than other models.

\subsection{Dataset Splits}\label{ssec:data_splits}

We now explore how model performance can be affected depending upon the dataset splits.

\textbf{Random Splits:} The performance of machine learning models is known to vary over different dataset splits. However, many benchmark datasets are provided with predetermined train/test splits, which has resulted, in the case of graph-based models, in approaches over-fitting to validation sets and not generalising to other random splits~\cite{shchur2018pitfalls}.

\begin{figure}[!ht]
    \centering
    \begin{subfigure}[b]{0.45\textwidth}
        \centering
        \includegraphics[width=0.99\textwidth]{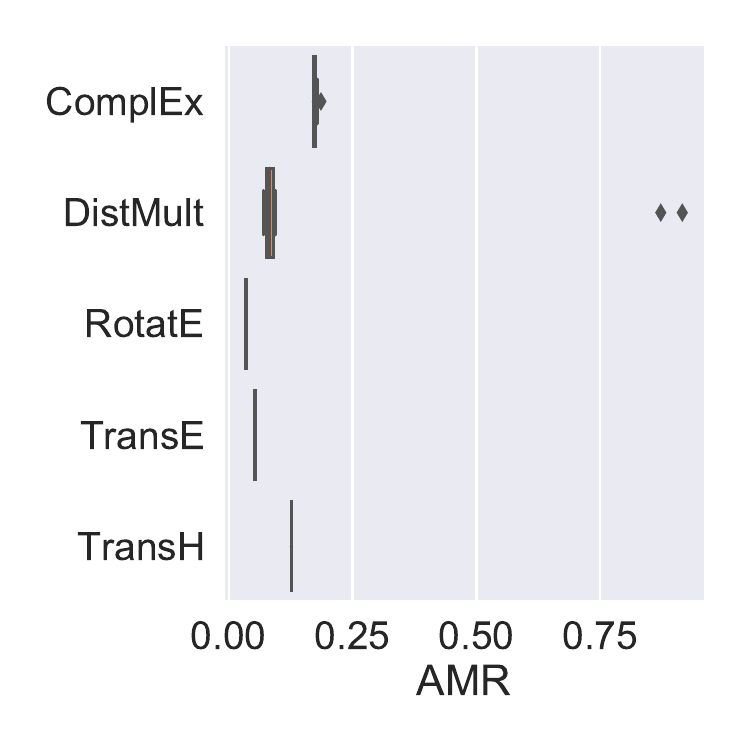}
        \caption{Hetionet}\label{fig:ds:hetnet}
    \end{subfigure}
    \begin{subfigure}[b]{0.45\textwidth}
        \centering
        \includegraphics[width=0.99\textwidth]{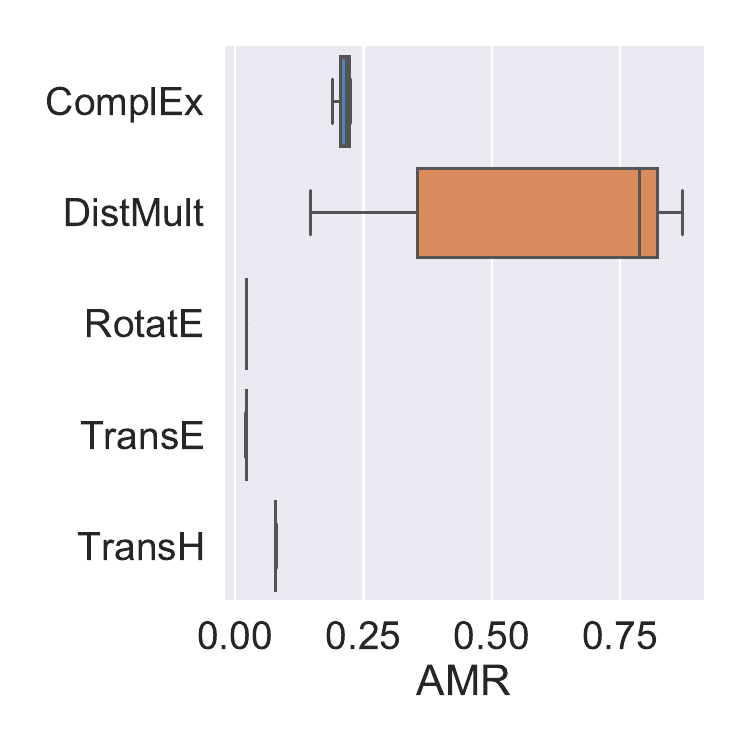}
        \caption{BioKG}\label{fig:ds:biokg}
    \end{subfigure}

    \caption{Distribution of AMR score obtained by repeating the experiment whilst varying the data split randomly.}\label{fig:dataset-seed}

\end{figure}

In this experiment, we assess how models respond to changes in the train/test split of the underlying datasets as other variables (hyperparameters, models seeds and training setup) are kept constant, with each dataset being split randomly 10 times. Here 10\% of the triples are used for test, with the remainder used for training. Table~\ref{tab:dataset-seed} displays the results from this experiment and shows that most models have very consistent performance across the different dataset splits. Indeed many of the values are similar to those from the model seed experiments, indicating that they do reflect the models true performance. However, Figure~\ref{fig:dataset-seed} shows that once again, that DistMult has much greater performance variability over the dataset splits which affects its average ranking compared with other approaches. This reinforces the notion that models should be tested both on different dataset splits and with different random initialisations to assess their true generalisability.

\begin{table}[h!]
    \tiny
    \centering

    \begin{tabular}{l l  c c  c c c  }
        \toprule
        \textbf{Dataset}          & \textbf{Approach} & \multicolumn{4}{c}{\textbf{Metric}} \T\B                                                                                        \\
        \midrule \midrule
                                  &                   & AMR \(\downarrow\)                       & MRR \(\uparrow\)           & Hits@1 \(\uparrow\)        & Hits@10 \(\uparrow\)\B     \\
        \cline{3-6}

        \multirow{5}{*}{Hetionet} & ComplEx           & 0.174\(\pm\)0.004                        & 0.024\(\pm\)0.006          & 0.007\(\pm\)0.002          & 0.059\(\pm\)0.016\T        \\
                                  & DistMult          & 0.242\(\pm\)0.343                        & 0.028\(\pm\)0.016          & 0.008\(\pm\)0.006          & 0.061\(\pm\)0.036          \\
                                  & RotatE            & \textbf{0.035\(\pm\)0.000}               & \textbf{0.126\(\pm\)0.000} & \textbf{0.135\(\pm\)0.001} & \textbf{0.262\(\pm\)0.001} \\
                                  & TransE            & 0.053\(\pm\)0.000                        & 0.081\(\pm\)0.000          & 0.035\(\pm\)0.001          & 0.117\(\pm\)0.001          \\
                                  & TransH            & 0.126\(\pm\)0.000                        & 0.032\(\pm\)0.000          & 0.006\(\pm\)0.000          & 0.087\(\pm\)0.001\B        \\
        \midrule
        \multirow{5}{*}{BioKG}    & ComplEx           & 0.212\(\pm\)0.012                        & 0.008\(\pm\)0.001          & 0.003\(\pm\)0.001          & 0.009\(\pm\)0.004\T        \\
                                  & DistMult          & 0.627\(\pm\)0.303                        & 0.014\(\pm\)0.003          & 0.007\(\pm\)0.001          & 0.026\(\pm\)0.004          \\
                                  & RotatE            & 0.022\(\pm\)0.000                        & \textbf{0.123\(\pm\)0.000} & \textbf{0.060\(\pm\)0.000} & \textbf{0.241\(\pm\)0.001} \\
                                  & TransE            & \textbf{0.021\(\pm\)0.000}               & 0.063\(\pm\)0.000          & 0.021\(\pm\)0.000          & 0.135\(\pm\)0.001          \\
                                  & TransH            & 0.079\(\pm\)0.001                        & 0.023\(\pm\)0.001          & 0.008\(\pm\)0.000          & 0.043\(\pm\)0.001\B        \\

        \bottomrule
    \end{tabular}
    \vspace{5pt}
    \caption{The mean performance with standard deviation over 10 fixed random dataset splits of all models and both datasets as measured by various metrics. Here only the random seed used to initialise the model parameters is changed, whilst all other variables are kept fixed.}\label{tab:dataset-seed}

\end{table}

\textbf{Domain Specific Splits - Gene-Disease Prioritisation:} One area where the use of knowledge graphs has great potential to be used within the drug discovery domain is gene-disease prioritisation~\cite{paliwal2020preclinical}. This crucial step in the drug discovery pipeline involves attempting to identify the causally implicated biological entity (typically a gene or protein) for a certain disease -- which can be thought of as predicting a link between a gene and disease entity in a knowledge graph. Drugs can then be developed to modulate this target entity and ultimately treat or cure the disease. As such, this task better reflects a real-world use case of knowledge graphs within the drug discovery domain.

In order to conduct this experiment, we focus upon the Hetionet dataset and evaluate performance at predicting links between gene and disease entities. Hetionet has three distinct relation types connecting genes and diseases: associates (12K examples), downregulates (7k examples) and upregulates (7k examples). For this work we, focus on the relation with the largest number of examples: associates. We partition the associates' edges using 10-fold cross validation, where 9 folds are used as part of the training data (along with the rest of the Hetionet dataset) and the remaining split used for solely test, with results presented as the average over all folds. One interesting aspect to consider is whether this could lead to trivial examples being present during training time, with which models could effectively cheat~\cite{rossi2020knowledge}. This is because if a gene and disease pair are linked via either a downregulates or an upregulates relation, they can also be linked via an associates relation (although this is only the case in a small number of pairs for the Hetionet dataset), meaning that the model could see gene and disease entity pairs both in the train and test set. To assess the impact of this, we create two versions of the training dataset, a biased version (where gene-disease pairs from the train set linked via other relations in the train set are not removed) and an unbiased version (where the training set contains no gene-disease pairs linked via other relations in the test set).

\begin{figure}[!ht]
    \centering
    \begin{subfigure}[b]{0.48\textwidth}
        \centering
        \includegraphics[width=0.99\textwidth]{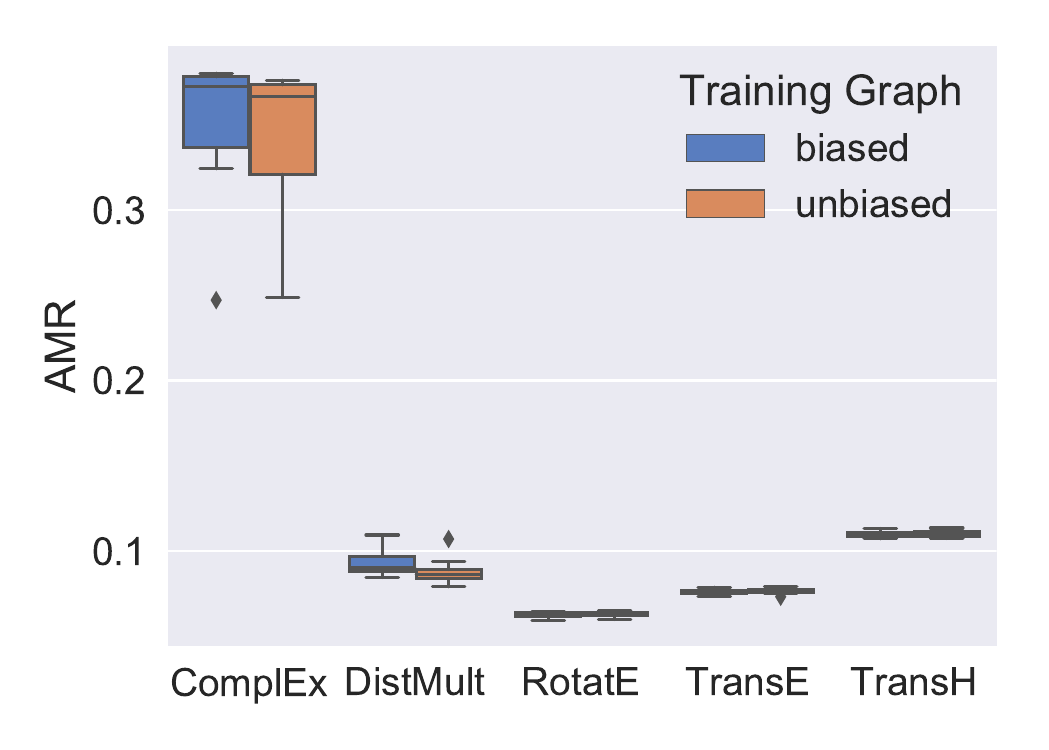}
        \caption{Bias Vs. Unbiased}\label{fig:gd:bias}
    \end{subfigure}
    \begin{subfigure}[b]{0.48\textwidth}
        \centering
        \includegraphics[width=0.99\textwidth]{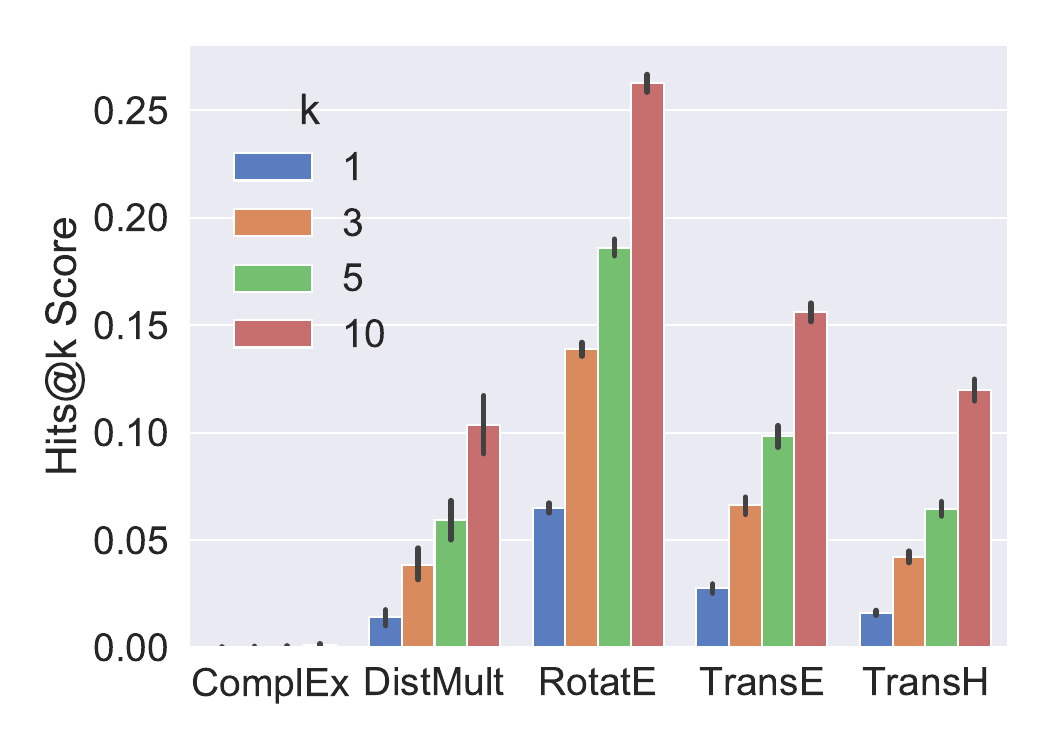}
        \caption{Hits@k}\label{fig:gd:hits}
    \end{subfigure}

    \caption{Assessing model performance at gene-disease prioritisation on Hetionet.}\label{fig:gene-disease-seed}

\end{figure}

Figure~\ref{fig:gd:bias} highlights the distribution of AMR scores on the test set over the 10-folds after training on the biased and unbiased datasets. Interestingly the figure shows that the potential bias introduced by trivial examples does little to impact performance in this instance. This could well be due to the small number of both gene-disease pairs and the even smaller number which are linked by more than one relation type -- however the wider issues of potential trivial examples in the context of drug discovery knowledge graphs is an area deserving continued attention.

Figure \ref{fig:gd:hits} shows the average Hits@k score over all folds using the unbiased data. One clear trend is that ComplEx performs poorly at the gene-disease prioritisation task, both compared to the other approaches and against its own performance in the general dataset split experiments. This may indicate ComplEx is more sensitive to low data quantity setups such as this. However, most models achieve performance on this gene-disease restricted setup which is comparable to that when run on the whole graph -- further reinforcing the notion that the crucial task of gene-disease prioritisation should continue to be investigated in the context of KGs.

\section{Discussion \& Conclusion}\label{sec:conclusion}

\textbf{Overall Observations.} Our results show RotatE is often the best performing model of the five on both datasets and throughout all the experiments. This reinforces previous similar findings~\cite{chang2020benchmark,ali2020bringing} and shows RotatE to be a strong baseline in the context of drug discovery. Our results also highlight that older approaches like TransE can still be very competitive given an optimised training and hyperparameter setup. Regarding training setup choices, we found that NSSA and AdaGrad were often the best performing approaches and could serve as starting points for further comparisons. More generally, it can seen that KGE models are more than just architectures and should be considered in combination with their training setup and hyperparameter values.

This study has further demonstrated the effect of hyperparameters, with tuning providing a potentially large increase in performance over default baseline choices. After performing a detailed parameter search using two sampling methods, it can be seen that there is still a large variance in parameter values, even among top performing configurations for a given model. This highlights how performance seems rarely to be driven by single parameters, rather a nuanced combination of values is often required. This may suggest hand-crafted tuning is unlikely to result in optimal choices and a HPO strategy should be employed. There was also a marked difference in hyperparameters between the two KGs, despite them being from the same domain and containing similar entities -- revealing how dataset dependent parameters can be. Additionally, we found a random search of parameter space (given enough repeats) to yield configurations which perform very closely to those from more principled approaches, whilst taking less time to do so.

We assessed model performance at target discovery by predicting links specifically between gene and disease entities. This showed that, although predictive performance was comparable to when measured on relations of all types, there were some differences. This suggests that researchers should not assume that model performance at the general link prediction task is indicative of performance in a more focused application. Additionally, we highlighted that performing HPO to optimise a single relation type can actually hurt downstream performance on that type in a limited data setting. The issue of potentially trivial examples being present deserves continued attention, especially in the context of drug discovery where the complexity of the underlying data could amplify the risks.

\textbf{Reproducibility \& Fair Comparison.} This work has shown how knowing model architecture alone, without training setup, hyperparameters and dataset splits, is probably insufficient to replicate results. To improve reproducibility, these should be more prominently reported alongside performance metrics. Additionally, to allow for fair comparisons to be made, new models should be assessed against well-tuned baseline approaches, with any changes introduced in training setup also being applied to competing methods. Also with the goal of fairer comparisons in mind, performance metrics should also be presented as averages over different random seeds and dataset splits in order to assess robustness.

\textbf{Conclusions.} KGs are increasingly being used in the field of drug discovery to help address key challenges such as gene-disease prioritisation. In this work we have assessed how various factors, including training setup, hyperparameter choices, model parameter initialisation and different dataset splits, affect the performance of five KGE models on two real-world biomedical datasets. However, ultimately for these approaches to impact drug discovery, they need to be used in associated decision making. This in turn depends on the level of trust and understanding of the approaches. We hope that increased attention on such foundational aspects will improve rigour, reproducibility and understanding of factors influencing different tasks and contexts, thereby maximising the potential to improve drug discovery efforts.

\textbf{Future Work.} We plan to expand the set of models we assess to include a more diverse set of approaches, as well as exploring other suitable drug discovery datasets and focusing on more domain specific tasks such as drug repurposing.

\section*{Acknowledgement}

The authors would like to thank Ufuk Kirik, Manasa Ramakrishna, Tomas Bastys, Elizaveta Semenova and Claus Bendtsen for help and feedback throughout the preparation of this manuscript. Additionally, we would like to thank all of the PyKEEN team, especially Max Berrendorf and Mehdi Ali for their help and support. We would also like to acknowledge the use of the Science Compute Platform (SCP) within AstraZeneca. Stephen Bonner is a fellow of the AstraZeneca postdoctoral program.

\bibliographystyle{plain}
\bibliography{RPbib}

\end{document}